# Multisensory Integration and Sensory Substitution Across Vision, Audition, and Hatpics: Answering the What, Which, and When in Study Protocols

Andrew Jeyathasan [1] and Swati Banerjee [1,*]

1    Université Paris-Saclay, CEA-LIST-91120, France
\*    Correspondence: andrew.jeyathasan@cea.fr (A.J.), swatibanerjee@ieee.org (S.B.), swati.banerjee@cea.fr (S.B.)

**Abstract:** We experience the world through multiple senses that work together to create a cohesive perception, whether in daily life or immersive technologies. Understanding this multisensory integration (MSI) requires examining the interactions between sensory modalities, each with unique temporal dynamics and characteristics. While most research focuses on unimodal or bimodal cues, the integration of three or more modalities remains underexplored. MSI studies must account for factors like cross-modal correspondence, congruence, cognitive load, and stimulus timing, which become increasingly complex as modalities multiply. This article examines these key factors and how they can be applied to design effective MSI study protocols.

**Keywords:** Multisensory Integration, Brain computer Interface, Immersive technologies; Cross-modal correspondence

## 1. Introduction

Multisensory integration (MSI) is a fundamental process that enables us to navigate and interact with our environment by combining information from various senses. Beyond basic perception, it plays a critical role in complex tasks (e.g. driving, playing a musical instrument) where interpreting visual, auditory, and tactile cues together is essential to perform efficiently the task [1]. This ability to merge inputs from different senses and combined into a unified experience, is at the core of immersive technologies, which aim to create a compelling sense of physical immersion [2].

Immersion is becoming increasingly popular, with significant advancements driven primarily by the desire to enhance video games experience, as the driving simulator [1]. An other example is the success of Pokemon GO, which blends augmented reality with real-world exploration creating a sense of immersion, but also through its game-play, which closely mirrors the classic Pokemon saga. By replicating familiar mechanics and story elements, Pokemon GO create the feeling to live in the Pokemon fictional world, enhancing the overall feeling of immersion [3]. These examples highlight two distinct forms of immersion, which is supported by Curran [2] works: mental immersion, as seen in Pokemon GO, where the game's narrative enhances the sense of being part of the Pokemon world, and physical immersion, shown by VR research which often discuss immersion in these physical terms [4]. To enhance this sense of physical immersion, such technologies often relies on familiar sensory experience. Research has shown that when simulations mirror real-life experience, they enhance immersion by allowing individuals to relate to the virtual world more easily [2]. In fact, Marucci et al. [1] talk about the benefit of naturalistic stimulus because of the reliability and time-locked activity in many brain regions compared to artificial and controlled stimuli [5]. However, the lack of controllability of naturalistic stimuli makes it hard to perform laboratory experience. To tackle this issue, we use innovative immersive technologies. One of the main immersive technology is Virtual Reality (VR) technology, particularly Head-Mounted Devices (HMDs), which plays a key role in creating immersive experiences. By fully engaging our visual perception and blocking out external distractions, these devices provide a complete visual immersion into the virtual world. We will talk more about immersive technologies in the Immersive technologies and BCI section. The study of MSI can also have a significant impact in healthcare. One of the main example where MSI is studied is in BCIs.

Brain-Computer Interfaces (BCIs) enable direct communication between the brain and external devices. Various BCI protocols have been designed based on different signal acquisition techniques, processing methods, and communication frameworks.

As highlighted in the review by Zhang, Zhou, and Jiang [6] and supported by the work of Ortiz-Rosario and Adeli [7], BCIs can broadly be classified into two categories: invasive and non-invasive. While invasive BCIs allow for more accurate recording of neural activity by placing electrodes directly into brain tissue, they require surgical procedures, making them less accessible and more difficult to study in typical research settings [8, 9].

Given these limitations, our interest lies in non-invasive BCI technologies, which are more suitable for widespread research and practical applications due to their safety and ease of use. Among non-invasive techniques, electroencephalography (EEG) stands out as one of the most accessible and extensively studied methods, with a vast body of literature supporting its use. EEG-based BCIs have shown promising applications in enabling communication for individuals with severe disabilities [10], in rehabilitation for stroke patients [11], and in neurofeedback therapies for conditions such as epilepsy, depression, and anxiety [12, 13, 14].

In their review of BCI spellers, Rezeika et al. [15] categorize EEG-based BCIs according to the paradigms they use. The most common approaches rely on Event-Related Potentials (ERPs)—time-locked electrical brain responses to specific sensory, motor, or cognitive events. ERPs differ from spontaneous EEG activity by their consistent amplitude changes following a stimulus, and their characteristics can vary depending on the nature and presentation of the stimulus. Among the most studied ERPs are the P300 and Steady-State Visual Evoked Potentials (SSVEP), which will be discussed further in the Assessment and Quantification of Multisensory Integration section [16]. Another widely used paradigm is Motor Imagery (MI), also referred to as Event-Related Desynchronization/Synchronization (ERD/ERS), which relies on changes in brain rhythms associated with real or imagined movement. These changes, particularly in the Sensorimotor Rhythm (SMR), are spatially localized based on the body part involved [16, 17].

BCI technology is advancing rapidly, both in terms of applications and underlying methods. However, the majority of BCI systems rely on visual ERPs to facilitate communication between the brain and external devices. As noted by Zhang, Zhou, and Jiang [6], this preference is largely due to the superior performance of visual ERP-based protocols across five key criteria: (1) target detection accuracy, (2) training time, (3) population adaptability, (4) information transfer rate (ITR), and (5) paradigm flexibility [18, 19]. This dominance of the visual modality can be attributed to the fact that primates (including humans) are inherently visual beings [20]. As a result, relying solely on other sensory modalities typically result in lower performance across these five dimensions, particularly in terms of population adaptability.

This reliance on visual modalities has created a technological gap in the development of non-visual BCI systems. Exploring non-visual BCIs is therefore an important area of research, particularly given the limitations of visual-only approache. As highlighted in the review by Wagner, Daly, and Väljamäe [21], non-visual BCI technologies are especially relevant for individuals with severe motor disabilities accompanied by visual impairments—such as patients with Amyotrophic Lateral Sclerosis (ALS), whose vision may deteriorate as the disease progresses. Furthermore, Zhang, Zhou, and Jiang [6] point out an additional drawback of visual-based BCIs: their reliance on the primary human sensory modality can significantly reduce a user's attention to their surrounding environment. This issue is particularly critical in practical applications such as intelligent wheelchairs, where maintaining environmental awareness is essential for safety and usability.

This is where MSI becomes crucial in BCI. Relying solely on other sensory modalities, such as auditory or tactile, typically results in poorer BCI performance compared to using visual stimuli alone [6, 22]. To match the performance of visual-based BCIs, bimodal BCIs have been developed by combining two modalities. These bimodal systems have shown improved or same performance over visual-only BCIs. For example, Wang et al. [23] demonstrated that bimodal stimuli can achieve comparable signal to visual-only stimuli. Building on this principle, Zhang, Zhou, and Jiang [6] developed a gaze-independent BCI speller using location-congruent auditory-tactile stimuli. Notably, Zhang, Zhou, and Jiang [6] highlight the importance of auditory and tactile spatial stimuli outside the field of view, while those within the field of view are less critical. Furthermore, incorporating the visual modality in bimodal designs can help reduce the cognitive load on the visual system, allowing users to remain more aware of their environment.

In the same study (see Fig.7 [23]), also explored trimodal stimuli that combined visual, auditory, and tactile modalities. Although trimodal stimulation did not significantly outperform bimodal stimulation, it showed slightly improved reaction times, P300 latencies, and P300 amplitudes. Beyond potential signal enhancement, trimodal stimuli may offer greater flexibility in BCI usage. For instance, Thurlings et al. [24] demonstrated that users could switch their attended modality in a visual-tactile BCI without compromising performance. This flexibility is promising, as it suggests that users could adaptively select their preferred modality, an approach that could be further enhanced through trimodal stimulation. This flexibility also offers a potential solution to the challenge that, outside of controlled experimental settings, certain modalities may be affected by environmental noise. Indeed, Marucci et al. [1] showed that multimodal signals are more robust and pronounced than unimodal signals in the presence of noise interfering with the stimuli.

Despite the potential of haptic, auditory and visual (HAV) cues to enhance performance and flexibility in BCI systems, there remains a significant gap in the literature regarding their use in such applications. To explore this untapped potential, we want to design a series of experiments aimed at assessing the impact of key parameters on MSI. The following sections outline the approach that can be taken to design these experiments, including the rationale behind parameter selection and task choice. Designing effective protocols first requires identifying reliable methods to induce and quantify MSI. This process not only guide the construction of MSI protocols but also laid the groundwork for acquiring insights into the mechanisms of MSI.

In this review article, we will primarily focus on three sensory modalities: vision, audition, and tactile. Particular emphasis will be placed on the use of EEG to record brain activity, as it provides valuable insights into the neural mechanisms underlying multisensory integration.

## 2. Inducing and Modulating Multisensory Integration

The human brain is an intricate network of billions of neurons communicating through complex pathways. This communication is essential for cognition, movement, sensory perception, and overall neurological function. Past research related to tactile-visual substitu-tion systems has explored the conversion of visual stimuli into tactile sensations, therefore laying the groundwork for sensory augmentation in assistive technologies [25, 26] to overcome the sensory overload in the visual cortex. Studies by Collins [27] have demonstrated the feasibility of inferring visual characteristics from tactile stimuli, underscoring the potential for cross-modal information transfer. This energy conversion phenomenon, where light energy is translated into mechanical or electrical energy, forms the basis of tactile-visual substitution systems [28]. Leveraging this principle, researchers have integrated electrotactile feedback into myoelectric hand prostheses, offering users enhanced sensory perception and control [29, 30, 31]. Such advancements promises to improve prosthetic devices' functionality and usability, empowering users with more intuitive and immersive sensory experiences. Oh, Yoon, and Park [32] presented a study to prove that localized electrotactile feedback outperforms visual feedback in a Virtual-reality-based table tennis game. However this approach remains to be very narrow in terms of applicability. In the following sections, we will explore how different modality can be linked together to create this cross-modal information transfer.

*2.1. Cross Modal Correspondence*

When two or more sensory modalities are presented together, they can influence each other in complex and meaningful ways. Each sense captures different aspects of the environment, yet these modalities can often convey similar or complementary information. This overlap suggests that the effectiveness of MSI depends on how similar or complementary the information is across senses, we then talk about Cross-Modal Correspondences (CMC). Cross Modal correspondence refers to the tendency for normal observers to match distinct features across different sensory modalities [33]. For instance, the association between a sound pitch and size of the object producing the sound [34]. Such correspondences are thought to be supported by shared or interacting brain regions that integrate multisensory input, allowing the brain to form a coherent representation of the environment [35].

CMC across the presented stimuli is essential for effective multisensory integration. Without proper CMC, the brain may struggle to combine inputs from different modalities, which can result in increased reaction times. This effect is illustrated in the pilot study [36] (see Fig.1), where congruent simultaneous cues led to longer reaction times compared to congruent sequential cues. In sequential condition, the visual cue was presented first, followed by the auditory and tactile cues 500 ms later. However, participants often responded before the onset of the auditory and tactile cues, effectively reacting only to the visual cue. Although the sequential condition was technically trimodal, it functioned as a visual unimodal cue in practice. So, using trimodal simultaneous cues leads to an increase in reaction time compared to a visual unimodal cue. This is a clear example of sensory overload, where the brain is unable to process all the information presented simultaneously, leading to higher reaction times and decreased performance.

Different types of CMC can be used to induce MSI such as location, temporal, semantic, familiarity, etc. The following sections will explore the most common CMC types studied to build our MSI protocol.

5 of 34...

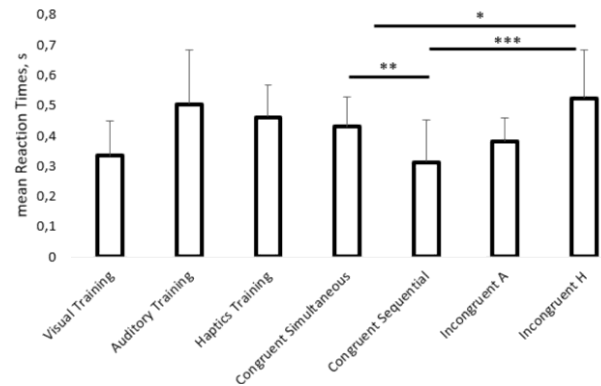

**Figure 1.** Mean reaction times and accuracies in experiment 1 across different tasks, *p¡0.05, **p¡0.01, ***p¡0.001 in paired t-test. Outliers were excluded from the reaction times. A - auditory, H - haptics. Adapted from Shumkova, Sabir, and Banerjee [36].

**Location-based correspondence**

There is different possibilities to incorporate MSI using various types of stimulus. The most commonly used correspondence in multisensory integration is location-based stimuli. Since vision, hearing, and touch all convey important spatial information that we frequently associate in daily life, the brain can more easily interpret multisensory stimuli when they are spatially aligned [37]. For example, a visual flash and a sound presented from the same position are more likely to be perceived as a single event [38].

**Temporal alignment**

Temporal alignment, or the simultaneity of multimodal stimuli is important for effective multisensory integration [39]. For instance, a tactile stimulus on the hand and a visual stimulus on a screen are more likely to be perceived as related when they occur at almost the same time [38], a phenomenon referred to as the Temporal Ventriloquist Effect.

To address this, the concept of the Temporal Binding Window (TBW) has been introduced—referring to the time interval within which stimuli from different modalities are likely to be integrated as a single perceptual event. Within this window, small temporal discrepancies between modalities can be tolerated, allowing for successful multisensory fusion.

**Relative unisensory strength**

Miller et al. [39] highlight the importance of relative unisensory strength in multisensory integration. For example, when a visual stimulus is paired with a much weaker auditory stimulus, the visual modality may dominate perception, resulting in a stronger response in the visual cortex compared to the auditory cortex. This imbalance can impact the multisensory enhancement effect, ultimately reducing the nonlinear interaction between modalities (see Fig.2)

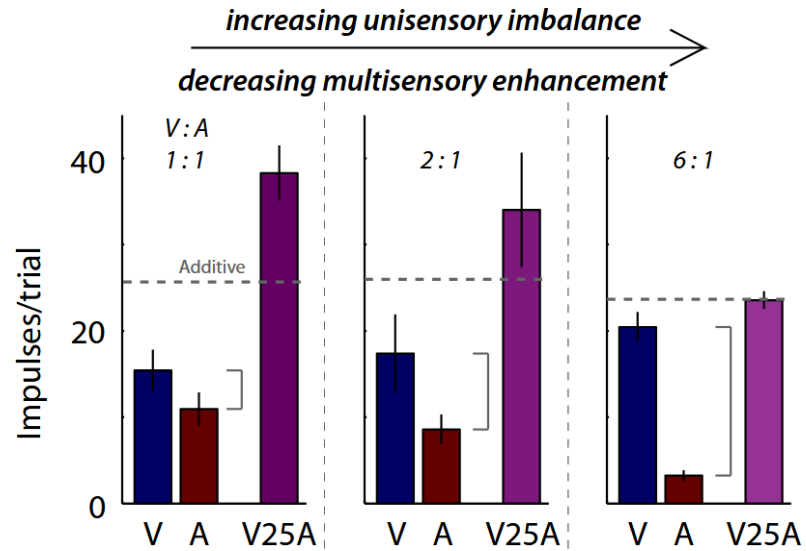

**Figure 2.** Adapted from Miller et al. [39]. Three exemplar neurons illustrate the trend in which the level of imbalance in a neuron's unisensory visual and auditory responses predicts the relative magnitude of its multisensory response to their combination. In each example, the sum of visual and auditory responses is approximately equivalent (horizontal dashed lines labeled "additive"), but as the unisensory response imbalance (shown at the top of each series of bars) grows, multisensory response magnitude decreases.

**Semantic congruence**

While location and temporal alignment are a common feature that is shared across all three modalities, we can also find associations of different features provided by different modalities. These associations are formed through our experiences interacting with the environment. For instance, a picture of a dog and the sound of dog barking is more easily associated due to their congruent semantic content [40, 41]. Stimuli that evoke similar emotional responses across different modalities can be perceived as more congruent. For example, a happy visual stimulus paired with a happy auditory stimulus can be perceived as more congruent than a happy visual stimulus paired with a sad auditory stimulus [42]. This shows that familiarity with a given association of stimuli can be perceived as more congruent [33].

**Other CMC features**

Features like the shape, frequency of the stimuli can be perceived as more congruent. For instance, a visual stimulus with a specific shape can be paired with a tactile stimulus that has a similar shape, enhancing the perceived congruence between the stimuli. [36, 43]

**Combination of multiple CMC**

It is also possible to combine multiple stimulus features to enhance brain responses, making it easier for the brain to integrate incoming information. Spatiotemporal features of stimuli are commonly exploited in various MSI applications [6, 41, 44, 24] as those aspects can be easily controlled and reproduced. The work of Pires et al. [41] demonstrates the effectiveness of using visuo-auditory stimuli with semantic, temporal, and spatial congruence (Figure 3) in a P300-based BCI for healthy patients. This BCI uses seven different words to facilitate communication of basic needs, which are selected through the oddball paradigm (see Section 3.1.1.1) induced by sensory stimulation and attention. However as it is highly based on attention and cognitive process, the result obtain with ALS patient in a completely locked-in state are not really promising.

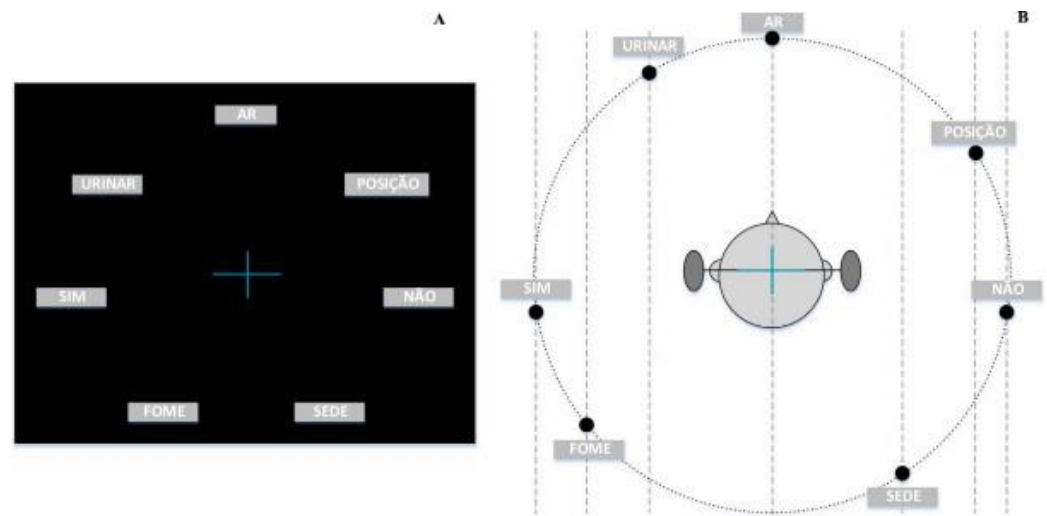

**Figure 3.** (A) Spatial location of the visual stimuli and (B) of the auditory stimuli (view above the head). Matching between visual layout and spatial sound location to achieve spatial congruence. [41]

*2.2. Workload*

Workload can significantly influence MSI. For instance, Marucci et al. [1] demonstrated that as workload increase through task difficulty increases, the brain tends to rely more heavily on multisensory input to maintain effective performance. In their study, difficulty was increased by introducing task-irrelevant sensory noise, mimicking real-life sensory challenges. Under these conditions, participants appeared to depend more on the integration of information from multiple senses, highlighting the brain's adaptive use of multisensory cues when faced with complex or noisy environments. Consequently, high-load tasks tend to accentuate the differences between unimodal and multimodal cues, both in behavioral performance and brain activity. This has important implications for BCIs in real-world applications, where environmental noise and distractions are common. In these contexts, leveraging multimodal cues can support the brain's natural reliance on multisensory integration, potentially improving BCI performance under challenging conditions.

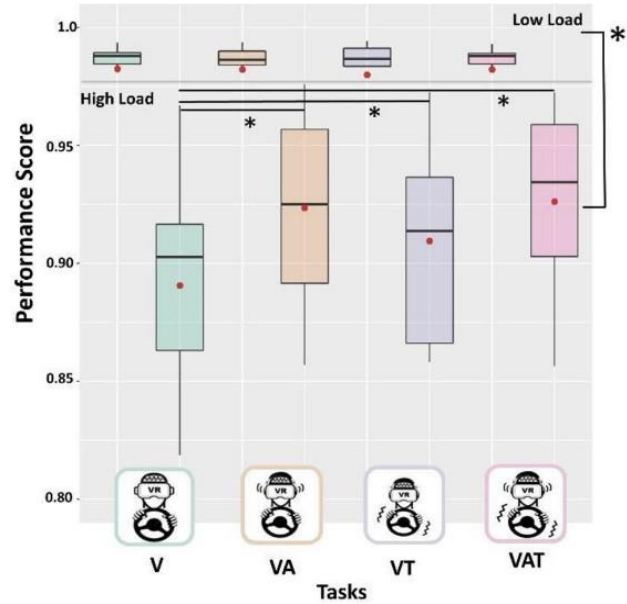

**Figure 4.** Adapted from Marucci et al. [1]. Boxplots repesenting behavioral correlates of the performance index for the low load condition (top) and high load (bottom) condition for each sensory task.

## 3. Assessment and Quantification of Multisensory Integration

Before delving into the experimental design, it is essential to establish a clear understanding of how to assess and quantify MSI. This section will outline the various methods and metrics used to evaluate MSI, including both behavioral and neural response measures. This understanding will serve as a foundation for the subsequent tasks' choice and experimental design.

### 3.1. Neurophysiological metrics

MSI has been assessed through a wide range of brain responses, spanning from large-scale brain networks to the activity of single neurons [35].

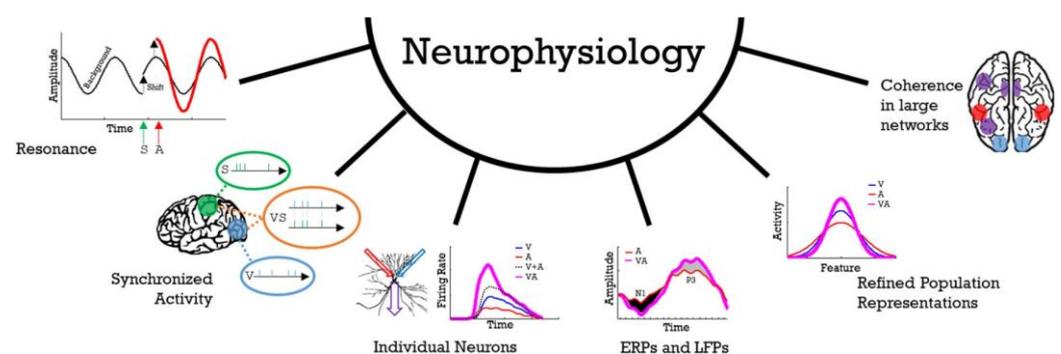

**Figure 5.** Adapted from Stein, Stanford, and Rowland [35]

#### 3.1.1. Event-related potentials (ERPs)

As presented earlier in the introduction, Event-related potentials (ERPs) are crucial in BCIs as they reveal brain activity patterns triggered by stimuli or cognitive processes. [45]

In medical contexts, ERPs are identified based on the direction of their electrical signals. If the signal moves upward, it is called positive (P), and if it moves downward, it is called negative (N). These labels are followed by a number that shows how many milliseconds it takes for the signal to reach its peak. ERP responses are unique to each sensory modality, with distinct patterns such as P1 and N1 emerging in different brain regions depending on whether the stimulus is visual, auditory, or somatosensory (Visual Evoked Potential (VEP), Auditory Evoked Potential (AEP) and Somatosensory Evoked Potential (SEP)) [22]. Different evoked potentials can be observed depending on how a specific sensory modality is stimulated, such as flash evoked potentials, checkerboard evoked potentials, and motion evoked potentials [15].

**P300 potential**

The most widely studied ERP is the P300 potential. It occurs at 300 ms after stimulus demonstration and is usually a positive deflection of voltage in the EEG trace. There is multiple hypothesis underlying the cognitive process inducing P300 : Novelty and surprise [46], context update [47], context closure [48], decision making [49], etc.

The procedure commonly used to study the P300 wave is the oddball paradigm. At least two different stimuli are presented to the subject, one being the non-target item (which appears frequently), and the other being the target item (whose appearance is rare and requires a response from the subject). There can also be a distractor item (which is infrequent and interferes with the target item). The fact that the target item is rare and requires a response informs us about the P3b, while the novelty provided by the rarity of the distractor item informs us about the P3a [50]. In complement of the oddball paradigm, particpants are usually asked to perform a running memory task. Participants are presented with a sequence of stimuli and must maintain and update a running count or sum of the stimuli. The P300 ERP is elicited as participants update their working memory with each new stimulus [51]. This task can also induce a Miss Match Negativity (MMN), a specif ERP induced when there is a discontinuity in a repetitive stimuli [52].

A task that resemble a bit the Oddball paradigm is the discrimination task. A traditional discrimination task in psychology typically involves presenting participants with stimuli and asking them to make a judgment or decision based on the differences between those stimuli. The goal of such tasks is to assess the ability to perceive and distinguish between different stimuli.

There are also other ways that are less common that are used to induce P300, as the tasks presented below activate neural processes associated with attention and memory retrieval. The Von Restorff effect, also known as the "isolation effect", predicts that when multiple homogeneous stimuli are presented, the stimulus that differs from the rest is more likely to be remembered [53]. The Sternberg task involves presenting a set of digits to a subject, followed by a brief delay. After the delay, a single digit (the probe) is presented, and the subject must quickly decide whether this digit was part of the original set [52]. There is also Task and rule switching task that can elicit P300, because the participant is required to update the context or mental model of the task/rule [54, 55].

Although early studies reported differences between visual and auditory P300 responses [56], later research challenged these findings. Today, the prevailing view is that P300 is largely modality-independent and likely arises from a shared neural generator [57]. Since all three sensory modalities (visual, auditory, and haptic) are thought to activate a shared neural generator, analyzing the P300 response can serve as a reliable means to assess and quantify multisensory processing. Figure 6 provides examples of P300 recorded for the different single modality-based stimuli.

Multimodal P300 responses have been shown to exhibit lower latency and a significant increase in amplitude, which likely contributes to faster processing of sensory information [23]. As illustrated in Figure 6, the P300 waveforms for different modalities and their combinations highlight these effects. We would like to replicate similar results with our experiment. In fact, with those results we will be able to observe brain responses that reflect the impact of MSI on integration time, and to demonstrate how HAV cues can enhance BCI performance by increasing P300 amplitude and reducing its latency.

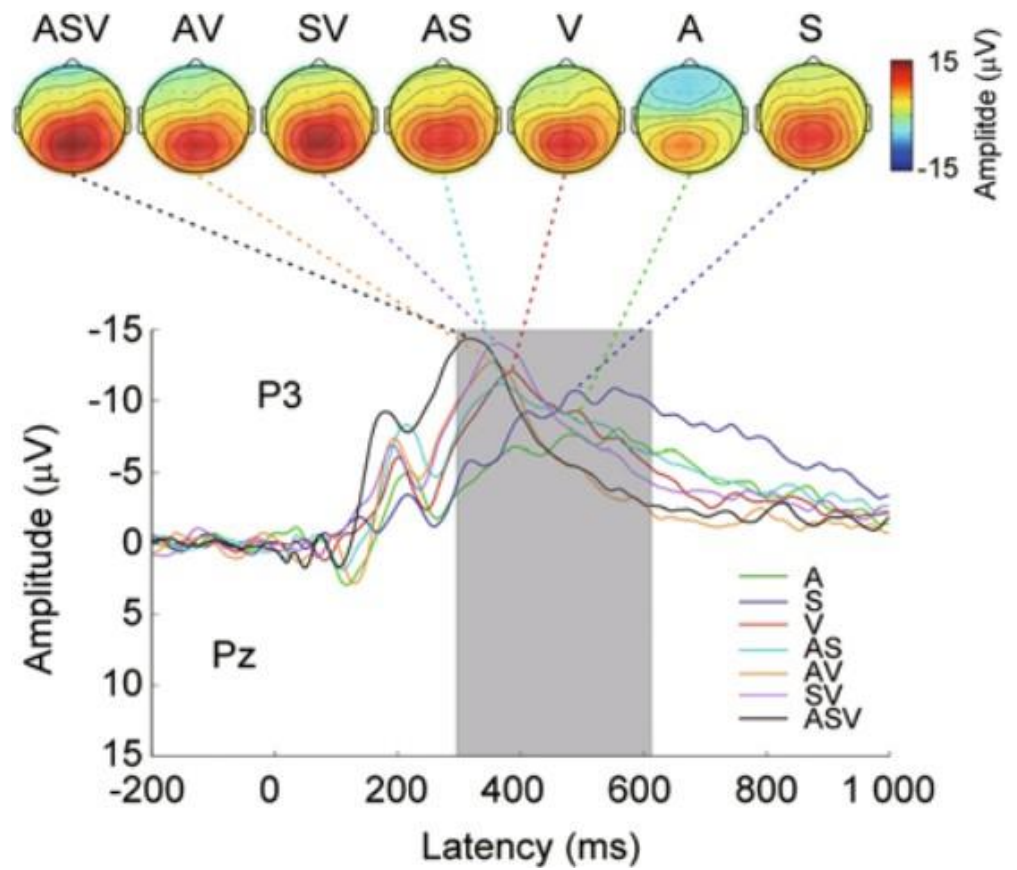

**Figure 6.** Adapted from Wang et al. [23]. A - auditory, S - Somatosensory, V - visual.

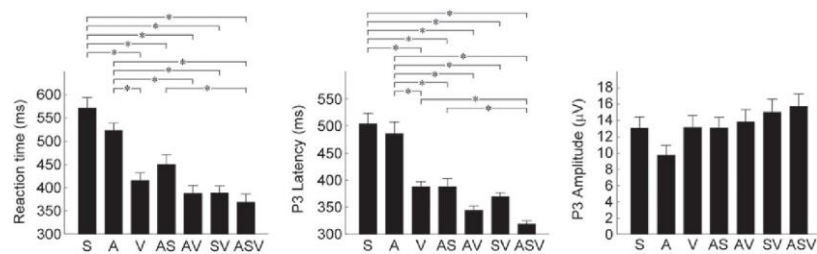

**Figure 7.** Comparison of reaction time (left panel), P3 latency (middle), and P3 amplitude (right) among the seven target stimuli. Mean ± SE. *P <0.05, one-way ANOVA [23]

**Steady-State Evoked Potentials (SSEP)**

Steady-State Evoked Potentials (SSEP) is a brain pattern that appears upon the stimulus presentation and can be used in BCIs as an alternative to P300 to increase the ITR [58]. Unlike P300 SSEP are relatively constant and periodic, occurring on the duration of longer time period. Moreover, SSEP is evoked by continuous or repetitive stimulation, while P300 is elicited even after a single short stimulus.

Muller-Putz et al. [59] designed a proof-of-concept study, in which they showed that vibrotactile-based SSEP from a single finger stimulation could be used in BCI [59]. Later by providing a vibrotactile stimulus to the 2 fingers and recording from larger set of EEG electrodes Breitwieser et al. [60] achieved similar results [60]. They also showed that BCI users responded to tactile modality with a broad tuning curve with maxima in a range from 21 to 35 Hz. Through combination of various EEG brain responses one can build a hybrid BCI. In another study vibrotactile SSEP were analyzed together with another pattern, called event-related desynchronization, and it was reported that hybrid BCI improved user's performance [61]. The research on visual-based SSEP was much more substantial. If the visual cue is presented at frequency above 10 Hz, the visual SSEP demonstrates an oscillatory behavior [62]. The frequency of stimulation is directly associated with the elicited SSEP response frequency. Until recent visual SSEP were mainly used to study the visual brain processing. However, several scientists have already tried using these brain patterns in the BCI applications [63]. It was shown that the BCI detection performance depends on the stimulation frequency. Several visual SSEP BCI spellers were developed achieving ITRs of 20-40 bits/min [64]. Hybrid visual SSEP and visual P300 BCI speller achieved an ITR of 56.44 bits/min showing a promising direction for BCI improvement through hybrid patterns. Finally, auditory SSEP are usually elicited at stimulation frequency of 40 Hz and occur mainly in the frontal and central scalp locations [65].

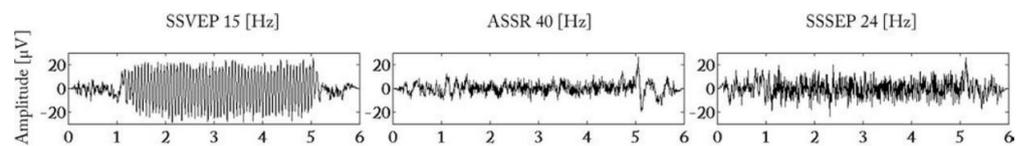

**Figure 8.** SSEPs from left to right for unimodal visual, ausitory and tactile stimuli at specific frequency. Adapted from Kus´ et al. [65]

Combination of several sensory modalities may potentially enhance the SSEP response and aid the BCI performance in a similar way to multimodal P300 through MI. With this rationale Kus´ et al. [65] investigated the SSEP at simultaneous delivery of visual, auditory and vibrotactile stimuli [65]. First, they observed the difference in topography of the unimodal SSEP. The strongest visual SSEP was observed at V1 and V5 areas. Auditory SSEP was the most prominent at the cenral electrode sites, while tactile SSEP resulted in the enhanced activity at the left parietal lobe site. When visual and auditory modalities were delivered simultaneously the SSEP magnitude was elevated in the occipital lobe and central areas. For the simultaneous visual-tactile stimuli the same effect was observed for the SSEP in the right parietal area and occipital lobe. Together these results provide no clear evidence of MSI happening in elicited bimodal SSEP. There is a necessity of further formal studies on the effect of the multimodal stimulation on the BCI performance as this is understudied due to various complexities associated to it.

3.1.2. Oscillatory response

Brain resonance refers to the phenomenon where inputs from one sensory modality (S) can influence the phase of ongoing neural oscillations, aligning them with an incoming signal from another modality (A) (see Fig.5). When the background oscillations resonate with the incoming signal, neural activity becomes more synchronized, leading to improved perception and encoding of the stimulus. In Atteveldt et al. [66] work, we see the importance of visual cues from the speaker in speech perception and processing, in fact there is the phenomena of phase resetting of oscillations in the auditory cortex induce by visual cues such as eyebrow raises, head inclinations, etc.

When inputs from multiple sensory modalities converge, they create a more reliable and distinct pattern of activation, reducing variability and noise in neural responses. This refined representation allows for more precise encoding of sensory information. For instance, studies on multisensory integration have shown that combining visual and vestibular cues in tasks such as heading discrimination leads to improved performance and reduced behavioral thresholds [67].

When multiple brain regions process congruent sensory inputs, their activity becomes more synchronized, leading to more stable and efficient neural communication. This coherence ensures that distributed populations of neurons work together more effectively, reinforcing the transmission of relevant information while minimizing interference from irrelevant signals. For instance, the study by Kumar et al. [68], found that during synchronous audio-visual speech perception, there was enhanced global gamma-band coherence, indicating increased synchronization among brain regions.

By looking at the certain spectral feature, it is also possible to extract information about workload using EEG. To select the subjective discriminant EEG spectral features related to the workload, we use a linear classification algorithm (automatic stop StepWise Linear Discriminant Analysis—asSWLDA) [69].This can be an interesting metrics as workload is thought to play a critical role in modulating MSI (see Workload Section).

*3.2. Reaction Time andInformation Transferring Rate (ITR)*

Reaction time (RT) is a widely used behavioral measure in cognitive psychology and neuroscience. It refers to the time taken by an individual to respond to a stimulus, typically measured from the onset of the stimulus to the initiation of a response. RT is often used as an indicator of cognitive processing speed and can provide insights into the efficiency of information processing in the brain. We determined that RT should be a key measure in our experiments, as it provides valuable insights into the efficiency of cognitive processing during multisensory integration. Evidence from a pilot study on the integration of HAV cues [36] demonstrated that RT was significantly influenced by both the congruency and timing of the presented stimuli (see Figure 1). These findings highlight the importance of considering RT as a sensitive indicator of how different sensory cues are integrated in the brain.

The importance of measuring RT was not only because of the cognitive performance. Banerjee and Jirsa [22] also highlight the significance of RT in the context of measuring BCI performance through the ITR. The ITR is a measure of the amount of information that can be transmitted per unit of time, and it is often expressed in bits per minute or bits per second. It is calculated using the equation:

$$ITR = B \cdot M \tag{1}$$

where $B$ is the number of bits per trial and $M$ is the mean number of decisions per minute. More information on the exact calculation of the ITR can be found in Serby, Yom-Tov, and Inbar [70] work. As an initial hypothesis, we can assume a direct relationship between ITR and RT, drawing an analogy in which the number of bits per trial ($B$) is represented by the reaction time in a human system. While this approach provides a useful starting point for exploring the connection between RT and ITR, it is important to recognize that this is a simplification. The relationship between RT and ITR is likely to be more complex and influenced by a variety of factors. To be able to use this hypothesis, we need to follow the frame drops during the experiment. The frame drop is the delay between two frames that are 10% greater than the screen refreshing period (which is usually 1/60 s). This can be caused by a variety of factors, including hardware limitations, software bugs, or network issues. Frame drops can lead to delays in stimulus presentation and affect the accuracy of RT measurements. To ensure accurate measurement of RT, we need to monitor variations in the screen refresh rate during the experiment and take appropriate measures to minimize its impact on the results. We implemented a code to monitor variations in the screen refresh rate, allowing us to identify and account for any potential issues that could explain outliers in the RT data and explore ITR (see Fig.9).

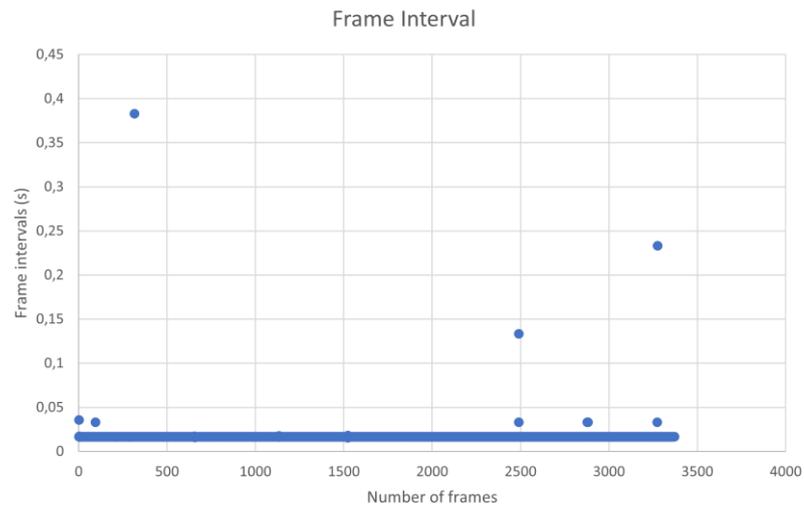

**Figure 9.** Frame interval during the Go/NoGo task visual staircase described in the section 6.1.2

*3.3. Subjective ratings*

Subjective ratings are a valuable tool for assessing and quantifying multisensory integration, for example in evaluating the sense of presence. The sense of presence refers to the feeling of "being there" in an environment, whether real or virtual, and is a critical metric in multisensory research. By examining how the sense of presence changes with the addition of sensory modalities, researchers can gain insights into the role of multisensory integration in creating immersive experiences. For instance, increasing the number of sensory modalities, such as combining visual, auditory, and tactile stimuli, often enhances the sense of presence [1]. Additionally, by manipulating factors like CMC, congruence, and timing of stimuli (see Section 2.1), we can look at the impact of those factors on the sense of presence. It can also be interesting to explore how immersive technologies (see Section 4), influence the sense of presence in comparison to more classical tools used to induce stimuli. By comparing the outcomes of similar tasks performed using immersive technologies versus traditional methods, researchers can gain insights into the added value of immersion and its impact on multisensory integration. In the driving test from Marucci et al. [1] work, they used a presence questionaire adapted from Wiederhold et al. [71] work, to assess the sense of presence. The questionnaire consisted of three questions scored on a scale from 1 to 7, which was the average across the three question to have a final presence score.

As developed in the Workload Section, assessing workload can be a valuable metric to evaluate the impact of multisensory integration. This can be done through a questionnaire, such as NASA-TLX [72], which provides a comprehensive assessment of workload across multiple dimensions. The NASA-TLX questionnaire consists of six subscales: mental demand, physical demand, temporal demand, performance, effort, and frustration. Participants rate each subscale on a scale from 0 to 100, and the scores are then averaged to obtain an overall workload score. This approach allows researchers to quantify the impact of multisensory integration on cognitive load and its relationship with performance in various tasks.

## 4. Immersive technologies and BCI

In the human perceptual system, all senses interact to create a unified experience of the multisensory environment. Immersive technologies, such as Virtual Reality (VR) and spatial audio, aim to replicate real-world sensory experiences, enhancing the feeling of full immersion. To achieve naturalistic stimuli, we have to create stimuli that is close to the real world experience. However according to Curran [2], having almost perfect real world experience can have the opposite effect and detract from the immersive effect because we have a hard time differencing the real and the virtual world. In fact, he say that an individual must make an effort to to be immersed, and that this immersion is a learned technique.

*4.1. Vision*

VR and Head Mounted Displays (HMD) are immersive technologies because of the feeling of being inside the virtual world, creating a convincing illusion of presence. This is possible because of HMDs' wide field of view, often up to 110 degrees, enhancing the sense of immersion. Real-time head tracking allows the virtual environment to adjust naturally, while interactivity through controllers makes users feel actively involved. The combination of visual and head tracking creates spatial awareness, leading to a strong sense of "presence" in the virtual environment [73].

In Lecuyer et al. [74] study, they showcase the usage of BCIs in VR application through video games, and serious games to improve BCI learning. They could perform navigation task in the virtual world, evaluate drivers' alertness level, having an immersive feedback of brain activity, solely using brain response like P300, SSVEP, Motor imagery... While this article emphasize the usage of BCI to improve communication in VR, we can also think the other way around where we use VR to induce interesting and immersive visual stimuli. As this study from Abbasi-Asl, Keshavarzi, and Chan [75], we can see that the usage of VR improve BCIs accuracy in a tasks consisting of object rotation or scaling in VR or 2D screen using either mental commands or facial expression (smile and eyebrow movement).

Augmented Reality (AR) technology is a powerful tool for inducing visual stimuli due to its ability to superimpose virtual images onto the real world [76]. This technology is becoming common, particularly in navigation systems, where it enhances spatial awareness and user interaction [77]. However, a major limitation of AR is its relatively small field of view, which restricts the spatial extent of visual stimuli and can limit the user's perception of their surroundings [78]. To overcome this constraint, some bimodal AR systems integrate auditory cues alongside visual cues, helping to direct the user's attention beyond the visible area and improve overall spatial orientation. This bimodal approach seems promising but it is limited by the Ventriloquism effect [76] (see Section 5.3).

This technology seems really promising, however Cornelio, Velasco, and Obrist [79] raise some limitations from this technology which is mainly size and distance misperception through VR [80] and cybersickness [81]. To tackle those issues, Cornelio, Velasco, and Obrist [79] presented the Particle-Based volumetric Display (PBD) technology which shows 3D images in mid-air [82]. So there is no need for head-mounted displays while giving 3D cues in the real world, enabling the possibility to perform tasks like depth discrimination [83]. There is also a MSI flexibility by using visual content that can be heard and felt simultaneously through a particular class of PDBs, Multimodal acoustic trap display (MATD) [84].

*4.2. Auditive*

To enhance the immersive quality of auditory experiences, controlling sound spatialization is essential. However, achieving this effect with physical speakers presents significant challenges due to their fixed positioning and acoustic constraints. An alternative approach is to simulate spatialization on stereo sound system using the Head-Related Transfer Function (HRTF), which replicates the way sound naturally interacts with the listener's ears and head [85]. However, HRTFs differ greatly from person to person, so to improve the spatial audio simulation, we can adopt a personalized HRTF method [86].

In [79] works, he talk about the utilization of Acoustic lenses and metamaterials which enables to control the directionality of sound, which can be used for spatial audio [87]. With this technology, it is possible to make levitate small bead in the air, but also provide a tactile sensation in the user's finger [88], enabling the possibility to do tasks like ventriloquism task (more info in Section 5.3).

*4.3. Somatosensory*

To stimulate the somatosensory pathway, the most currently used technology are through deformable surfaces, force feedback, data gloves, or vibration actuators on skin [79]. Those stimuli are mostly used in size estimation or identification tasks.

Cornelio, Velasco, and Obrist [79] introduce a technology that enables tactile sensations in midair through ultrasonic phased arrays. This approach has been employed in recent studies to replicate classic multisensory paradigms, such as the rubber hand illusion and the apparent tactile motion effect. These findings suggest that non-physical tactile stimuli can still generate sensory inputs comparable to those produced by direct physical contact. This technology offer sone multisensory flexibility by combining mid-air tactile and auditory stimulation simultaneously. This also can be an addition to virtual and augmented reality technology to provide multisensory integration.

## 5. Sensory Integration protocols

There is numerous experimental protocols that have been developed to study sensory processing in all three modalities (visual, auditory, and tactile). There is tasks that are designed to elicit specific ERPs or oscillatory responses, which can be used to assess sensory processing and multisensory integration (see Section 3.1.1).

*5.1. Unimodal experiments*

The most common ERPs used in BCI are the P300 [15]. Different tasks are designed to elicit the P300 response, such as the Oddball paradigm, Go-NoGo (GNG) task, and Two-Alternative Forced Choice (2AFC) task. In the GNG task, participants encounter a sequence of stimuli and must selectively respond to a designated "Go" stimulus while withholding responses to a "NoGo" stimulus. They are asked to give a response to the Go stimulus to asses the reaction time. A 2AFC task is a behavioral paradigm in which a participant is presented with a stimulus and must choose between two possible options, even if unsure. For example, they might decide whether a visual or auditory signal appeared on the left or right. The key feature of 2AFC is that a response is always required, forcing a decision between the two alternatives. This setup allows researchers to measure perceptual sensitivity and decision-making while minimizing response bias.

Where as GNG and 2AFC tasks requires a response from the participant and more speseeically motor response, the BCI speller task does not require a response from the participant, but rather relies on the detection of specific brain activity patterns using the oddball paradigm. The oddball paradigm is a widely used experimental protocol designed to study attention, sensory perception, and cognitive processing [89, 90]. In this task, participants are presented with a sequence of repetitive "standard" stimuli interspersed with infrequent "deviant" stimuli. By manipulating the frequency, complexity, or modality of the deviant stimuli, researchers can explore how sensory systems adapt to changes and prioritize task-relevant information [91]. This protocol is used mainly in the application of BCI spellers [15]. In a visual P300-based speller, the oddball paradigm is used by flashing rows and columns of a character grid in random order. The user focuses attention on the character they wish to select, and only the row and column containing that target character are meaningful to them. Since these target-containing flashes occur rarely compared to all other non-target flashes, they act as "oddball" stimuli. When an oddball (i.e., a target row or column) appears, it elicits a distinct P300 wave in the user's EEG, which the system detects and uses to identify the desired character.

Steady-State Evoked Potential (SSEP) tasks offer a distinct approach to studying sensory processing compared to P300 ERP. Unlike these tasks, which often rely on discrete stimuli and participant responses, SSEP tasks involve the presentation of continuous, repetitive stimuli at specific frequencies, allowing researchers to measure the brain's steady-state responses directly. This method provides unique insights into the temporal dynamics of sensory processing, as it enables the tracking of neural activity in real-time and the assessment of how sensory systems synchronize with external stimuli. An SSVEP-based BCI speller uses steady-state visual evoked potentials (SSVEP) to allow users to select characters by focusing on flickering visual targets. Each character or symbol on the screen flickers at a unique frequency, and when the user fixates on one, their visual cortex generates brain activity at the same frequency. This frequency-specific response is detected in the EEG signal and decoded in real time to determine which character the user is attending to. SSVEP-based spellers are known for their high speed and accuracy, making them efficient for communication in BCI applications. Because motor response is not required, the BCI speller task can be used to assess sensory processing without the confounding effects of motor activity.

While the previous tasks are known to elicit specific ERPs, we can also look at tasks that evoke a specific oscillation in the brain. Visual perception tasks, such as motion detection and dynamic pattern recognition, is a good example. This paradigm is used to study how the brain processes moving or changing visual stimuli. For instance, tasks involving the detection of coherent motion in a field of moving dots or tracking shifting patterns are known to evoke gamma-band (30–80 Hz) oscillations, which are associated with neural synchronization and information integration in the visual system [92]. A growing area of interest is the potential to adapt these tasks to other sensory modalities, such as auditory or tactile domains. For example, auditory motion detection tasks could involve identifying changes in the direction of a moving sound source, while tactile tasks might require detecting vibrations or patterns moving across the skin [93].

All of these tasks can be adapted to study MSI by combining unimodal protocols into multimodal versions, as each protocol is compatible with visual, auditory, and tactile modalities.

### 5.2. Artificial MSI

Combining unimodal protocols into multimodal versions is a powerful approach to studying MSI as it is based on well established and robust unimodal tasks that have been extensively validated. By adapting existing unimodal tasks, researchers can explore how the brain processes and integrates information from multiple sensory modalities. This method allows for the investigation of how different sensory inputs interact, influence each other, and contribute to overall perception and cognition. For example, in the traditional GNG task, participants respond to a single modality, such as vision or audition. By introducing multisensory stimuli, such as pairing a visual "Go" signal with an auditory and/or tactile cue, researchers can explore how the integration of information from different senses influences response accuracy, integration time and inhibition [56].

However, it is essential to consider how different sensory modalities are associated within a task. As discussed in the Inducing and Modulating Multisensory Integration section, various types of CMCs can be used to bind sensory inputs : semantic, spatial, temporal, and motion-based. The congruency of these CMCs can significantly influence how the brain integrates sensory information. For instance, in a GNG task, semantic CMCs can be employed to enhance multisensory integration (see Section 6.1.1)

In Misselhorn et al. [94] work, they presented a Trimodal Congruency Judgment task based on a 2AFC task, to evaluate the impact of sensory congruence using trimodal stimuli (see Section 6.3). Compared to the GNG task, this task incorporates a temporal dynamic component, as participants must evaluate the congruency of the stimuli over time rather than at a single moment. This task can be further adapted to incorporate various types of CMC. For example, motion-based CMC can be implemented by pairing moving visual stimuli (such as a moving object) with auditory cues that vary in pitch (e.g., a rising or falling tone) (see Fig.10 [95]), and by introducing dynamic tactile stimuli using devices like a haptic tablet.

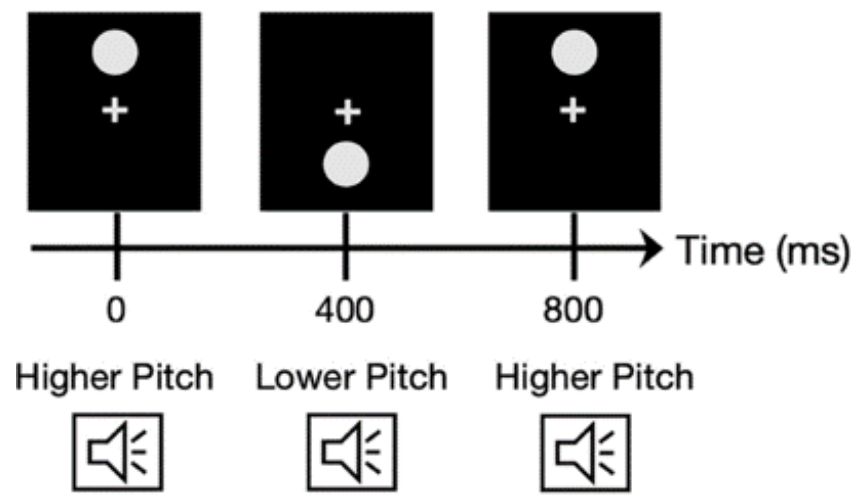

**Figure 10.** Schematic illustration of Visual elevation and auditory pitch CMC. Adapted from Uno and Yokosawa [95].

As discussed in Section **Relative unisensory strength**, achieving a balance between the different sensory modalities is essential for effective multisensory integration. To ensure this balance, we implemented a pre-experiment calibration phase in which individual perception thresholds were determined for each modality. These thresholds correspond to the minimum intensity at which each participant can reliably detect the stimulus. By setting the stimulus intensity to a fixed percentage above these individualized thresholds, we created a balanced condition where the unisensory strength of each modality is equalized. This approach not only maximizes the potential for multisensory enhancement but also accounts for inter-individual variability, allowing the task to be tailored to each participant's perceptual capabilities.

Studying MSI using SSEPs using mutlimodal cues is particularly interesting because SSEPs provide a frequency-specific, continuous, and objective measure of sensory processing in the brain as in Nozaradan, Peretz, and Mouraux [96]. By tagging each sensory modality (e.g., visual and auditory) with distinct stimulation frequencies, researchers can track the brain's response to each input separately and simultaneously. This allows them to observe how the brain dynamically integrates inputs based on temporal features, like synchrony. In the context of MSI, this method reveals how temporal congruency enhances cortical responses and phase coherence, suggesting that the brain binds multisensory signals more efficiently when they are temporally aligned. SSEPs therefore offer a powerful window into the real-time neural dynamics of multisensory processing, with excellent signal-to-noise ratio and the ability to isolate modality-specific contributions to integration.

To explore the temporal congruency of multisensory stimuli, we can use the Simultaneity Judgment (SJ) and Temporal Order Judgment (TOJ) tasks [97]. In SJ tasks, participants are presented with pairs of stimuli from different modalities and asked to judge whether they were perceived as simultaneous or not. This task allows researchers to investigate the temporal binding window (TBW) of multisensory integration, which is the time frame within which stimuli are perceived as a single event. In TOJ tasks, participants are presented with two stimuli in rapid succession and must determine their order of occurrence. This task provides insights into how the brain processes temporal information and integrates sensory inputs across modalities.

Inducing MSI artificially by combining unimodal tasks provides researchers with precise control over the experimental stimuli. This level of control enables systematic manipulation of various perceptual features, such as volume, opacity, and amplitude, allowing for targeted investigation of how these parameters influence MSI. However, while these tasks often rely heavily on CMC to facilitate integration, they remain quite distinct from the naturalistic stimuli encountered in everyday life. In real-world settings, CMCs and MSI are shaped by rich, dynamic, and context-dependent sensory experiences, which are difficult to fully replicate in controlled laboratory environments. As a result, artificially induced MSI, though valuable for isolating specific mechanisms, may not capture the full complexity of multisensory processing as it occurs in natural contexts.

*5.3. Multisensory integration paradigms*

There are several paradigm to allows us to explore MSI effect. The Ventriloquist Effect is a classic example of how two sensory modalities can interact to influence perception. In this task, participants are presented with a visual stimulus (e.g., a moving object) and an auditory stimulus (e.g., a sound) that are spatially misaligned. Despite the discrepancy, participants often perceive the auditory stimulus as originating from the location of the visual stimulus, illustrating the brain's tendency to integrate information across modalities [98]. While the Ventriloquist Effect is commonly associated with the spatial alignment of stimuli, it can also be used to study their temporal alignment. For instance, the auditory stimulus may be perceived as occurring simultaneously with the visual stimulus, even when they are temporally misaligned [99]. Furthermore, this effect is not limited to audio-visual cues but can also occur with audio-tactile [100] and visuo-tactile cues [101], highlighting its relevance across multiple sensory modalities.

In addition to the Ventriloquist Effect, numerous other multisensory phenomena demonstrate how the brain integrates information across sensory modalities. Examples include the McGurk Effect, where conflicting auditory and visual speech cues alter the perception of spoken sounds [102], and the Audio-Tactile Loudness Illusion, in which auditory and tactile stimuli combine to influence the perceived intensity of a sound or vibration [103]. Similarly, the Rubber Hand Illusion highlights the integration of visual and tactile cues to create a sense of ownership over a fake hand [104]. However, these effects are typically restricted to interactions between two specific sensory modalities, such as audio-visual, audio-tactile, or visuo-tactile, whereas the Ventriloquist Effect can be used to study interactions across all three modalities. To extend these paradigms to include a third sensory modality, task-irrelevant stimuli can be introduced to increase cognitive load (see Section 2.2). In fact, increasing workload through the addition of task-irrelevant stimuli closely mirrors the complexity of real-world environments, making it one of the most naturalistic approaches for studying multisensory integration.

*5.4. Real-world like task*

What is particularly interesting about MSI paradigms is that they are based on lifelong multisensory integration, naturally developed through everyday experiences. As development progresses, the brain undergoes dynamic reweighting of stimulus characteristics and learned associations, which ultimately shapes multisensory processing, which results in extensive cross-modal plasticity. This reweighting process allows the brain to adapt to changing sensory environments and optimize the integration of information across modalities. For instance, early in development, sensory modalities may contribute equally to perception, but as individuals gain experience and form associations, certain modalities may become more dominant depending on the context or task demands [105]. An example of cross modal plasticity can be the Sound induced flash task where participants are presented with a series of visual flashes and auditory beeps, with some beeps not accompanied by a flash. In some cases, participants perceive a flash in response to the beep alone, demonstrating cross-modal plasticity in multisensory integration.

More immersive tasks, such as driving simulations, provide a naturalistic and ecologically valid way to study MSI. Driving is a highly multisensory activity that engages vision, audition, and tactile sensations. Visual cues are critical for navigation and obstacle detection, auditory cues provide information about the environment (e.g., honking or engine sounds), and tactile feedback, such as vibrations from the steering wheel or seat, can serve as task-irrelevant stimuli that still contribute to the overall sensory experience. The familiarity of driving for most participants enhances the ecological validity of the task, making it an ideal paradigm for studying MSI in real-world contexts [1]. Some studies have demonstrated that naturalistic signals, compared to artificial stimuli, can produce more reliable and consistent outcomes [1].

*5.5. Inter-Participant variability*

Individual differences play a significant role in how participants perform tasks, particularly in cognitively demanding paradigms like the GNG task. For instance, if the stimuli are presented too quickly, some participants may experience an increased cognitive load, leading to slower reaction times and higher error rates. To address this variability and ensure a similar cognitive load across participants, a staircase approach can be implemented. Using this staircase we can find the perceptual threshold of the participant for a specific feature (e.g., stimulus intensity, duration, or frequency). The staircase procedure can be used to find the TBW for the SJ and TOJ tasks (see Section 6.2.2) or to set the GNG task stimuli (see Section 6.1.2).

Additionally, MSI does not function uniformly across individuals. Factors such as age, sensory modality dominance, and prior experience can influence how effectively participants integrate information from multiple sensory modalities. For example, some individuals may rely more heavily on visual cues, while others may prioritize auditory or tactile inputs, leading to differences in task performance. Therefore, it is crucial to account for these individual differences when designing experiments to ensure that findings are not skewed by such biases and are representative of the broader population.

## 6. Examples of MSI tasks

*6.1. Go/NoGo task*

For the first task we designed a Go/NoGo (GNG) task. The GNG task, examines response inhibition and decision-making by requiring participants to respond to specific stimuli while withholding responses to others [106]. Because the GNG task is a well-established paradigm, we were confident that using it would yield meaningful brain responses and more specifically it will elicit P300. Additionally, its straightforward design makes it easier to manipulate stimulus parameters, allowing us to effectively modulate MSI across all three modalities [56, 107].

6.1.1. Stimulation

In this task, participants encounter a sequence of stimuli and must selectively respond to a designated "Go" stimulus while withholding responses to a "NoGo" stimulus.

The stimuli used in the GNG task were deliberately selected to reflect familiar experiences from everyday life, ensuring that participants could intuitively interpret each cue (see Section 2.1.0.4). Visual stimuli were presented on a screen, auditory cues were delivered through headphones, and haptic feedback was provided via a PS4 controller, which also served to record participant responses. For the Go condition, a green checkmark was displayed alongside a bright, smooth tone commonly associated with positive feedback or successful actions, and no vibration was given—mirroring the typical absence of haptic feedback during correct actions in video games. In contrast, the NoGo condition featured a red cross, a sharp, harsh sound linked to errors or negative outcomes, and a vibration from the controller, reflecting the haptic feedback often experienced when making mistakes in gaming contexts. By aligning the sensory cues with both semantic meaning and emotional valence, this design aimed to enhance multisensory integration and optimize task performance by leveraging natural, real-world associations.

6.1.2. Thresholding procedure

The staircase phase consisted of four staircase procedures (see Fig. 13). Each staircase was designed to calibrate a specific parameter for a sensory modality: the Go and NoGo visual stimuli opacity, the Go and NoGo auditory stimuli volume and the NoGo tactile stimuli frenquency (see Section A). Before each staircase, participants completed a short practice session using salient changes in the relevant modality to familiarize themselves with the task. After each staircase, a verification test was conducted using the individualized thresholds determined by the staircase.

All stimulus parameters used in the experiment could be set to values ranging from 0 to 1. For each staircase, the initial stimulus intensity was set near zero and progressively adjusted during the procedure until it reached the participant's individual threshold.

6.1.3. Experimental paradigm

The visual, auditory, and haptic Go/NoGo stimuli were then presented for a maximum of 500 ms and at 150% of the estimated threshold determined in the Thresholding procedure phase. Participants were instructed to press the X button on the controller in response to a Go stimulus, and to withhold their response for a NoGo stimulus (see Fig. 11).

The experimental phase consisted of seven blocks, each corresponding to a specific type of stimulus presentation: three unimodal (visual, auditory, and haptic) and three bimodal (any combination of two modalities: visual-auditory, visual-haptic, or auditory-haptic) and a trimodal block (all three modalities presented together). A stimulus was considered a Go trial if at least one of the presented modalities corresponded to a Go stimulus (see Fig.12). Each block consisted of 20% Go stimuli and 80% NoGo stimuli (see Fig.12). The order of the blocks was randomized for each participant.

A final block was included at the end, where unimodal, bimodal, and trimodal cues (visual, auditory, and haptic presented together) were presented in a randomized order. A stimulus was considered a Go trial if at least one of the presented modalities corresponded to a Go stimulus (see Fig.12). This block consisted of 20% Go stimuli (with equal proportions of the seven possible cues) and 80% NoGo stimuli (with equal proportions of the seven possible cues) (see Fig.12).

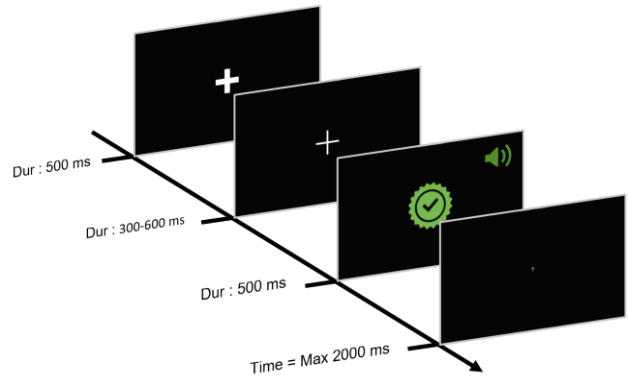

**Figure 11.** Overview of the GNG task experimental design. In this example of a trial, the participant is presented with a Bimodal Go stimulus, so the participant must respond by pressing the space bar.

6.1.4. Adaptive test phase

To increase the difficulty of the task, we designed an adaptive test phase where we reduce those features of the stimuli:

- Visual stimulus opacity: The stimuli become more transparent, making them less salient as they blend into the background.
- Auditory stimulus volume: The auditory stimuli become quieter by reducing the amplitude of the signal, making them harder to detect.
- Haptic stimulus frequency: The vibration frequency delivered through the controller is reduced, making the haptic feedback less salient and more difficult to perceive.

To do so, we use the same design as the last block of the Experimental paradigm phase where unimodal, bimodal, and trimodal cues are presented in a randomized order. This block is repeated 5 times with 20% Go stimuli (with equal proportions of the seven possible cues) and 80% NoGo stimuli (with equal proportions of the seven possible cues).

Through this phase, the stimuli become less salient, thereby increasing the difficulty of the task. To reduce the percentage of signal change, we adjust the threshold percentage (initially set to 150% during the Experimental paradigm phase). The threshold decreases from 138% to 90% in steps of 12% across the 5 blocks (see Fig. 13).

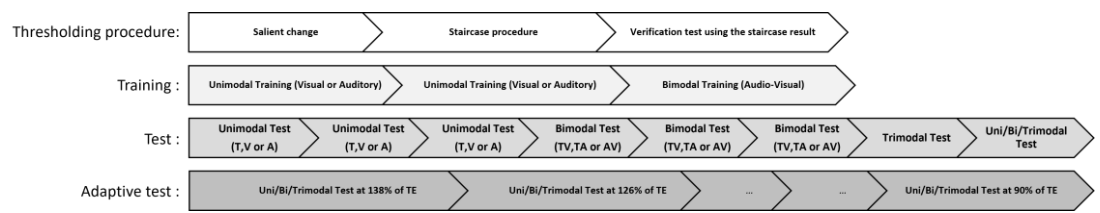

**Figure 12.** GNG task Test phase blocks and number of trials for each stimulus.

**Figure 13.** Overview of GNG task protocol

## 6.2. Perceptual judgment task

In the following task, we use a 2AFC task to investigate the effects of temporal alignment on MSI developed by Binder [97] (see Section 5.1). As for the Go/NoGo task, we were confident that using a 2AFC task would yield meaningful brain responses and more specifically it will elicit P300. In addition to the task design by Binder [97], we also incorporated an 6.2.4 phase to adapt the difficulty added EEG recording during the task.

### 6.2.1. Stimulation

To investigate the effects of temporal alignment on MSI, we incorporated a Perceptual Judgment task (PJ) developed by Binder [97] into our experimental battery. This task includes both a Simultaneity Judgment component (SJ) (where they must judge whether two stimuli are presented simultaneously or not) and a Temporal Order Judgment (TOJ) component (where they must judge the order of two stimuli presented in different modalities), designed to assess how temporal alignment influences multisensory processing.

For this task, we used a faded white circle as the visual stimulus and a 500 Hz tone as the auditory stimulus (see Fig.24(c) ). These two stimuli were selected based on the study by Binder [97], although they do not elicit any CMC. However, since our focus is on the effect of temporal alignment on multisensory integration, using two simple stimuli is sufficient to investigate this phenomenon.

### 6.2.2. Thresholding procedure

During the first phase, the threshold estimation phase (TE), the individual simultaneity thresholds were determined for each subject. Simultaneity threshold is the point on the psychometric curve where there is an equal probability of the 'simultaneous' and 'non-simultaneous' response. Thus, the range between simultaneity thresholds for 'sound-first' and 'flash-first' indicates the SOA values for which the audiovisual pairs are more often judged as synchronous than not (see Fig.17), refered as Temporal Binding Window (TBW). An interleaved staircase procedure was used to determine simultaneity thresholds, and it consisted of four staircases with the instruction to judge simultaneity of the presented audiovisual pairs. The SOA values listed below are expressed as the delay of the sound stimulus onset. The initial SOA values of the four staircases were -250 ms, 0 ms for the two staircases with the leading sound stimulus (sound-first trials), and 0 ms, 250 ms for the two staircases with the leading visual stimulus (flash-first trials) (see Fig.14).

The simultaneity thresholds were calculated separately for sound-first and flash-first trials, as means of SOA values of the last decisions in each of two staircases.

By determining each participant's individual TBW, this task enables us to compare behavioral and neural responses in conditions where stimuli fall within the TBW—allowing for the ventriloquist effect—and outside the TBW, where such integration is less likely to occur.

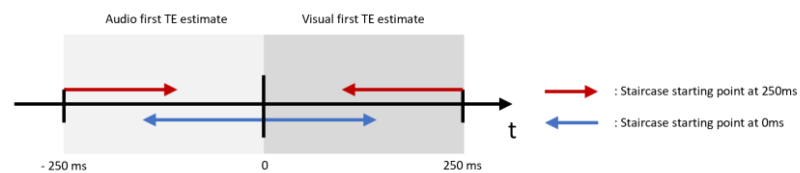

**Figure 14.** Illustration of the staircase procedure. The initial SOA values of the four staircases were -250 ms, 0 ms for the two staircases with the leading sound stimulus (sound-first trials), and 0 ms, 250 ms for the two staircases with the leading visual stimulus (flash-first trials).

6.2.3. Experimental paradigm

During the experimental phase, participants performed SJ and TOJ tasks in separate blocks, interleaved with rest periods. The order of the blocks was randomized. At the beginning of each trial, a warning stimulus was displayed—four gray squares surrounding the location of the upcoming visual stimulus. After a randomly selected delay, the auditory and visual stimuli were presented with a specific stimulus onset asynchrony (SOA). SOA values used during the S-T phase were based on the individual simultaneity thresholds determined during TE phase. There were six possible SOA values: equal to simultaneity threshold, half of the threshold value and twice of the simultaneity threshold value, both for sound-first and flash-first trials. Probability of each SOA value was identical and for each trial was 0.1667 (see Fig.17). Presentation of both stimuli terminated simultaneously, and then a message was displayed at the center of the screen to tell his decision ("Did the stimulus happen simultaneously (press "R2") or not (press "L2")" if it was SJ or "Which came first? Visual (press "R2") or auditory (press "L2") cue ?") and reminding which button of the controller will lead to which answer. The message disappeared when subject responded. There were 6 blocks with the SJ task and 6 blocks with the TOJ task 19. For each block, six experimental trials were presented, and the rest of the period duration between the consecutive blocks was 7 s. Before the beginning of each block a text message was displayed informing about the incoming task. It was associated with a warning sound. The block started 100 ms after the disappearance of the message. Moreover, the color of the warning stimuli displayed during each task block signaled the type of the task and in the center of the screen short text captions informed the subject about the current block (see Fig.15).

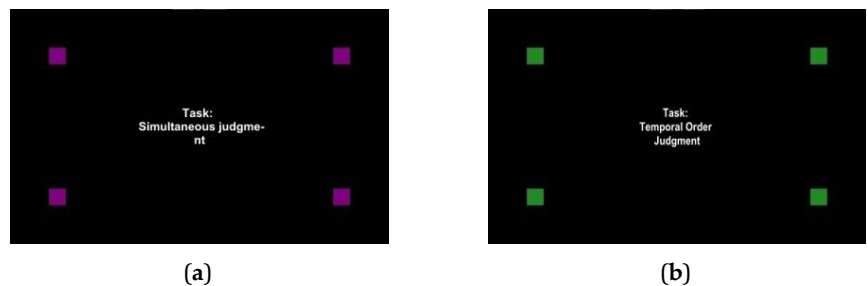

**Figure 15.** Screenshots of the warning stimuli for the Simultaneity Judgment (SJ) task (a) and the Temporal Order Judgment (TOJ) task (b).

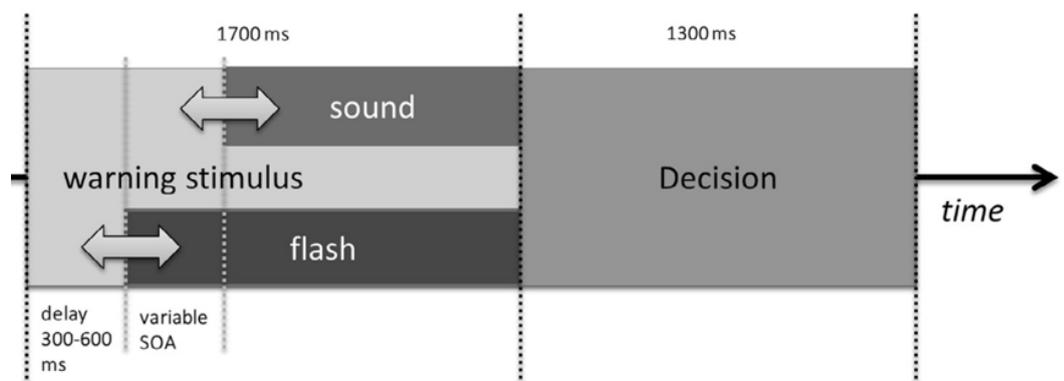

**Figure 16.** Illustration of the trial design during the PJ task. Adapted from Binder [97].

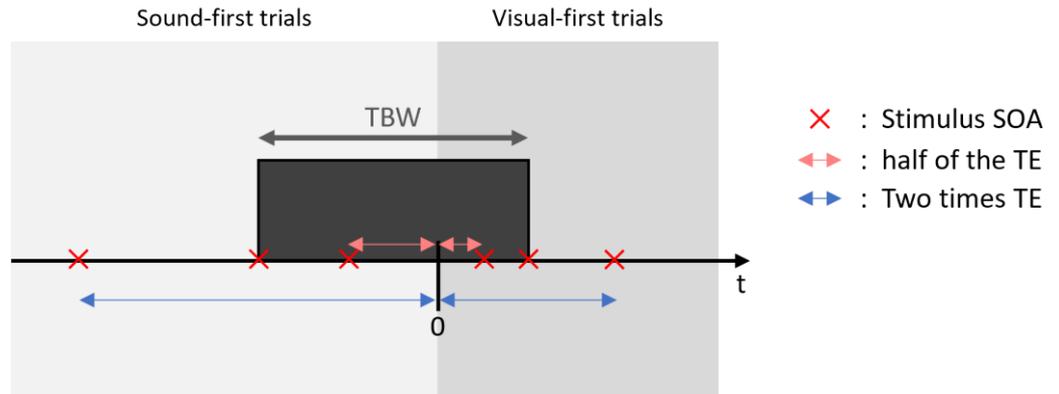

**Figure 17.** Illustration temporal location of the six SOA values.

6.2.4. Adaptive Test phase

In this phase, there will be 12 block (6 SJ and 6 TOJ) with 6 trials each using the same design as the Experimental paradigm but using varying SOA values 19. In fact, to increase the difficulty of the task, we decided to give to the participant, stimuli that are more and more close to the TE (see Fig. 18) so that it is harder for the participant increasing therefore his attention. To make it gradually more difficult, we decided to calculate the new SOA values as follows:

- n times the simultaneity threshold : $n * TE$
- One over n times the simultaneity threshold : $\frac{1}{n} * TE$

We use this formula both for sound-first and flash-first trials TE, with $n$ decreasing from 1.925 to 1.1 in steps of 0.15 every two blocks. This ensures we have recordings for every n values of both the SJ and TOJ tasks. The randomness between the SJ and TOJ tasks only determines which task comes first in each pair of blocks.

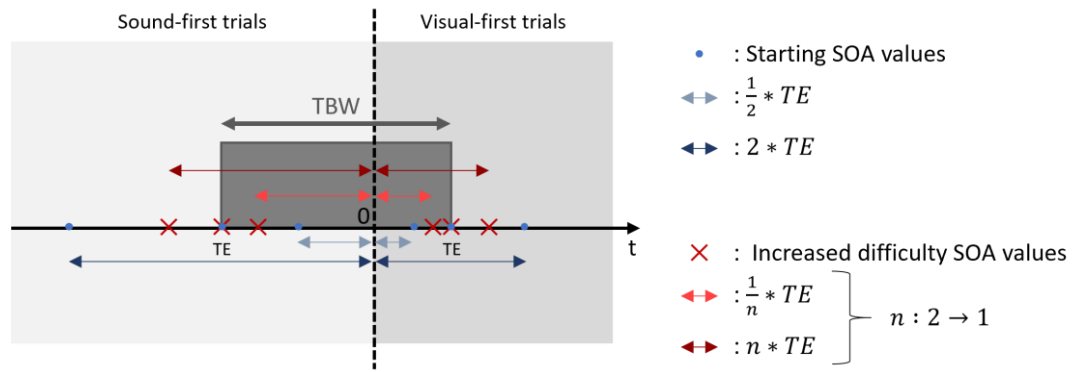

**Figure 18.** Illustration of the changing SOA during the PJ task adaptive test phase.

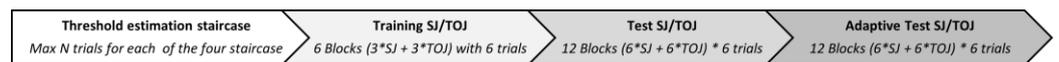

**Figure 19.** Overview of PJ task protocol

6.3. Congruency judgment task

To investigate the effects of congruency on MSI, we incorporated a Congruency Judgment (CJ) task into our experimental battery, based on the paradigm described by Misselhorn et al. [94]. The CJ task is also a 2AFC task like the Perceptual judgment task. However, here we are interested in the effects of congruency on MSI. In addition to the task design by Misselhorn et al. [94], we also incorporated an 6.3.4 phase to adapt the difficulty added EEG recording during the task.

6.3.1. Stimulation

The CJ task is designed to assess how congruency between different sensory modalities influences participants' responses. In this task, participants are presented with trimodal cues and instructed to attend two of the three modalities. They are then required to judge whether the attended stimuli are congruent (i.e., both increasing or both decreasing in signal amplitude) or incongruent (i.e., one increasing and the other decreasing).

In all trials, the three sensory modalities were stimulated with 2-second signals that underwent a transient change in amplitude. The visual stimulus consisted of a centrally presented circular grating presented on a monitor screen. The auditory stimulus was a 500 Hz sinusoidal tone presented through a headphones. Vibrotactile stimulation was implemented throught a PS4 controller (see Section A).

The CMC in those stimuli are primarily dynamic, as congruency is based on the temporal pattern of signal change—specifically, whether the amplitude in each modality increased or decreased over time.

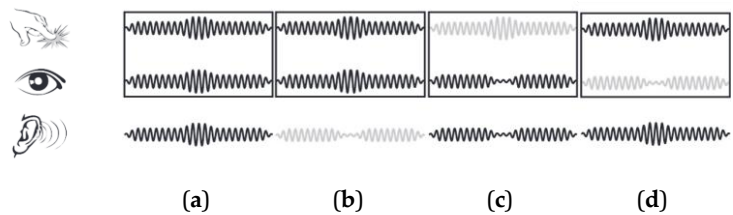

(a) (b) (c) (d)

**Figure 20.** Congruence patterns for the visual–tactile focus (analogous for audio–visual and audio–tactile foci). Cartoons depict tactile (first row), visual (second row) and auditory (last row) stimulation. Boxes indicate the attended combination of modalities. Congruently changing modalities are depicted by a black line and incongruent modalities by a gray line. While all stimuli are congruent in (a), one stimulus is deviant in (b),(c),(d), respectively. In (b), attended stimulus are congruent and the deviant is task irrelevant. For the two attended incongruent cases, the auditory distractor is either congruent to the attended visual stimulus ((c), audio–visual congruence) or to the attended tactile stimulus ((d), audio–tactile congruence). Adapted from Misselhorn et al. [94].

6.3.2. Thresholding procedure

The magnitude of change for a given modality and direction (increase or decrease) was individually adapted to ensure comparable perceptual salience and detection performance across modalities and change directions. The following procedure was implemented (Fig. 22(c)). Initially, salient unimodal sample trials of both changes per modality were presented to get the participants acquainted with the stimulus material. These trials were repeated until the participants reported to have an intuitive feeling for how these stimuli and, most importantly, their changes look, sound or feel. Subsequently, a threshold was estimated with a reversed staircase procedure. Initial change magnitude was close to zero and was incremented from trial to trial with a fixed step-size. This was repeated until participants reported to be sure to have perceived the change. Subsequently, the direction of changes with the previously estimated magnitudes had to be judged by the participants. If they were not able to correctly categorize 80% or more of the presented changes, estimated thresholds were increased, and the task was repeated. After that, the actual threshold estimation was carried out using an implementation of the Bayesian adaptive psychometric method QUEST. To this end, the two stimuli were presented concurrently with only one stimulus changing per block that had to be attended. The occurrence of increases and decreases of intensity was randomized and the magnitude of change was iteratively varied over 30 trials per change direction. The estimate from the preceding staircase procedure served as the initial prior. Subsequently, the last trial from the respective change direction in combination with information on success or failure to detect the change served as input for the Bayesian model to compute presentation intensity for the next trial. After 30 trials, stable estimates of detection thresholds were reached (see Fig 21).

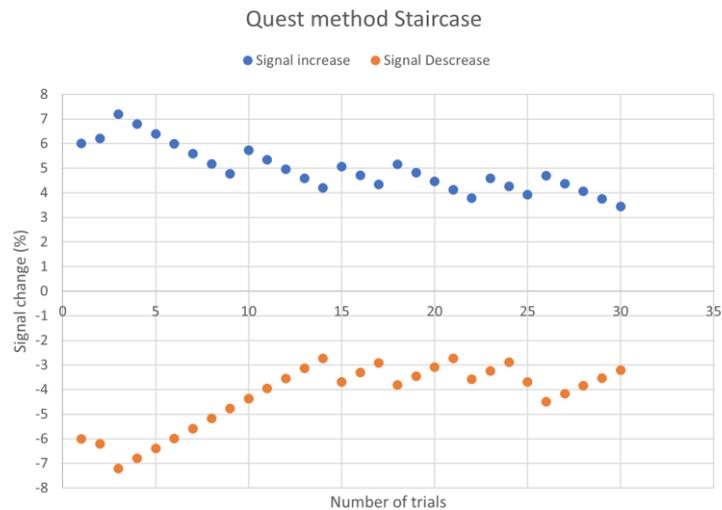

**Figure 21.** This graph represents the evolution of the TE (in %) during the thresholding procedure of CJ task.

6.3.3. Experimental paradigm

Concurrent visual, auditory, and somatosensory stimulation lasted for 700–1000 ms until a change in stimulus intensity occurred for 300 ms simultaneously in all modalities (ramping up: 100 ms, maximal change intensity: 100 ms, ramping down: 100 ms). After change offset, stimulation continued for another 700–1000 ms depending on the pre-change interval length (total duration of stimulation was always 2 s). After the presentation, participants respond using the controller (response interval max. 3000 ms). The participants' task was to evaluate congruence between changes of an attended pair of modalities irrespective of the change in the third modality. The attentional focus was cued block-wise as audio–visual (AV), visual–tactile (VT) or audio–tactile (AT). A change was defined as "congruent" if the changes in the two attended modalities had the same direction, e.g., an increase in contrast of the grating and a concurrent increase in amplitude of the vibration while VT was cued for this block (Fig 20(**a**),20(**b**)). The direction of change in the ignored modality was irrelevant to the decision. A change was defined as "incongruent" if the directions of change in the attended modalities differed, e.g., an increase in vibration intensity and a concurrent decrease in contrast of the grating while VT had been cued for this block (Fig 20(**c**),20(**d**)). After each trial, the participant was asked to report if the presented cues were congruent (by pressing "R2") or incongruent (by pressing "L2"). Each block consisted of 32 trials in random order and each stimulus configuration occurred 4 times. The order of attentional foci was randomized for each participant (e.g. AT, AV, VT) and repeated 2 times. The complete experiment comprised a total of 192 trials (6 blocks with each 32 trials) (see Fig. 23). For each subcondition (cells in Fig. 22(**b**)), 16 trials were recorded.

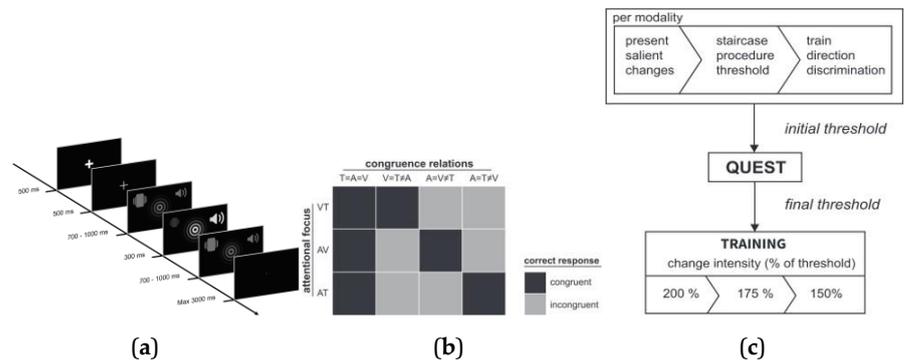

**Figure 22.** Overview of study design and procedure. In this example of a trial (a), visual contrast increases (central grating), vibration strength (T) decreases and auditory amplitude (A) increases. Correct answers (b) for all possible congruence relations ("=" is congruent, "≠" is incongruent) per attentional focus (VT=visual–tactile, AV=audiovisual, AT=audio–tactile). Thresholding and training procedure (c) preceding the experiment. Adapted from Misselhorn et al. [94].

6.3.4. Adaptive Test phase

In this phase, there will be 15 block (5 AV, 5 AT and 5 VT) with 32 trials each using the same design as the Experimental paradigm but using lower signal changes (see Fig. 23). In fact, to increase the difficulty of the task, we decided to give to the participants cues where the change in the signal is less salient, increasing in consequences the difficulty of the task. The decrease in signal change will happen gradually through the blocks and in all three modalities at the same time. The percentage of the threshold will go from 138% to 90% in steps of 12% every three blocks. This ensures we have recordings of every percentage value for the AV, AT and VT blocks. The randomness between the AV, AT and VT tasks only determines which task comes first in each set of three blocks.

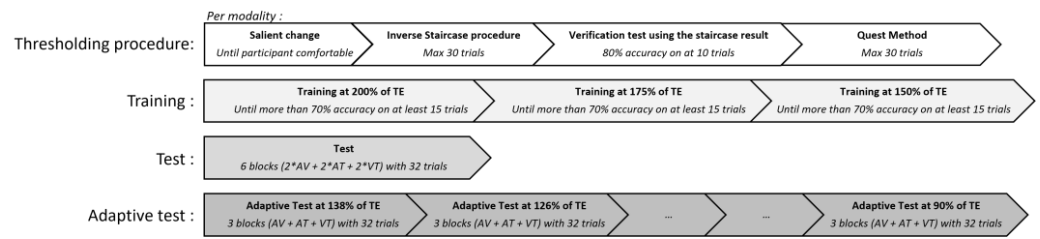

**Figure 23.** Overview of CJ task protocol

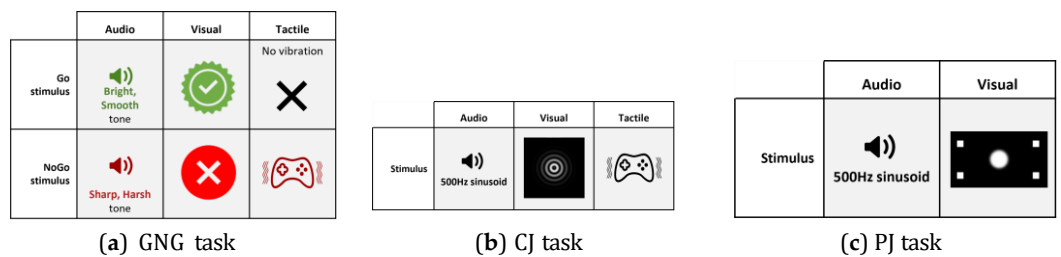

(**a**) GNG task  (**b**) CJ task  (**c**) PJ task

**Figure 24.** Stimuli used in the GNG, CJ, and PJ tasks.

## 7. Discussion

Authors should discuss the results and how they can be interpreted from the perspective of previous studies and of the working hypotheses. The findings and their implications should be discussed in the broadest context possible. Future research directions may also be highlighted.

## 8. Conclusions

This section is not mandatory, but can be added to the manuscript if the discussion is unusually long or complex.

## Appendix A : PS4 Controller Description

The PS4 controller used in this experiment is a widely available and familiar device, making it an ideal choice for participants who are accustomed to using game controllers in various contexts. Its buttons and triggers were employed to record participants' responses, enabling seamless integration into the experimental setup.

The PS4 controller (DualShock 4) is equipped with an integrated vibration feedback system that enhances user immersion by delivering tactile sensations. This system relies on two internal eccentric rotating mass (ERM) motors, one in each handle, which generate vibrations by spinning off-center weights at varying speeds. The two motors have different weights, allowing for distinct vibration intensities in each handle (see Figure A1). The intensity and frequency of the vibrations are modulated by adjusting the motor speeds via software, using a control variable ranging from 0 to 1. However, this approach does not allow for independent or direct control over vibration frequency and intensity. For more precise haptic feedback, a device that enables separate adjustment of these parameters would be preferable.

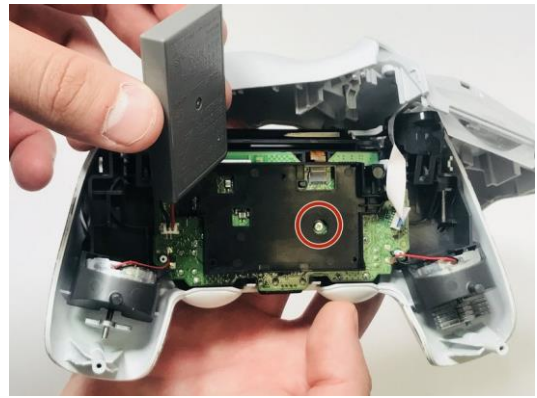

**Figure A1.** Teardown of the PS4 controller showing the internal components, including the vibration motors. The two motors are located in each handle of the controller, providing tactile feedback during gameplay.

## References

[1] Matteo Marucci et al. "The impact of multisensory integration and perceptual load in virtual reality settings on performance, workload and presence". en. In: *Scientific Reports* 11.1 (Mar. 2021), p. 4831. DOI: 10.1038/s41598-021-84196-8.

[2] Noirin Curran. "Factors of Immersion". en. In: *The Wiley Handbook of Human Computer Interaction*. Ed. by Kent L. Norman and Jurek Kirakowski. 1st ed. Wiley, Feb. 2018, pp. 239–254. DOI: 10.1002/9781118976005.ch13.

[3] Lili Liu, Christian Wagner, and Ayoung Suh. "Understanding the Success of Pokémon Go: Impact of Immersion on Players' Continuance Intention". en. In: *Augmented Cognition. Enhancing Cognition and Behavior in Complex Human Environments*. Ed. by Dylan D. Schmorrow and Cali M. Fidopiastis. Cham: Springer International Publishing, 2017, pp. 514–523. DOI: 10.1007/978-3-319-58625-0_37.

[4] Giuseppe Civitarese. "Immersion versus interactivity and analytic field". en. In: *The International Journal of Psychoanalysis* 89.2 (2008). _eprint: https://onlinelibrary.wiley.com/doi/pdf/10.1111/j.1745-8315.2008.00019.x, pp. 279–298. DOI: 10.1111/j.1745-8315.2008.00019.x.

[5] Uri Hasson, Rafael Malach, and David J. Heeger. "Reliability of cortical activity during natural stimulation". In: *Trends in cognitive sciences* 14.1 (Jan. 2010), p. 40. DOI: 10.1016/j.tics.2009.10.011.

[6] Boyang Zhang, Zongtan Zhou, and Jing Jiang. "A 36-Class Bimodal ERP Brain-Computer Interface Using Location-Congruent Auditory-Tactile Stimuli". eng. In: *Brain Sciences* 10.8 (Aug. 2020), p. 524. DOI: 10.3390/brainsci10080524.

[7] Alexis Ortiz-Rosario and Hojjat Adeli. "Brain-computer interface technologies: from signal to action". en. In: *Reviews in the Neurosciences* 24.5 (Oct. 2013). Publisher: De Gruyter, pp. 537–552. DOI: 10.1515/revneuro-2013-0032.

[8] Alexis Ortiz-Rosario et al. "Combined corticospinal and reticulospinal effects on upper limb muscles". In: *Neuroscience Letters* 561 (Feb. 2014), pp. 30–34. DOI: 10.1016/j.neulet.2013.12.043.

[9] M Spüler et al. "Decoding of motor intentions from epidural ECoG recordings in severely paralyzed chronic stroke patients". en. In: *Journal of Neural Engineering* 11.6 (Oct. 2014). Publisher: IOP Publishing, p. 066008. DOI: 10.1088/1741-2560/11/6/066008.

[10] Saba Moghimi et al. "A Review of EEG-Based Brain-Computer Interfaces as Access Pathways for Individuals with Severe Disabilities". In: *Assistive Technology* 25.2 (Apr. 2013). Publisher: Taylor & Francis _eprint: https://doi.org/10.1080/10400435.2012.723298, pp. 99–110. DOI: 10.1080/10400435.2012.723298.

[11] Xun Chen et al. "The Use of Multivariate EMD and CCA for Denoising Muscle Artifacts From Few-Channel EEG Recordings". In: *IEEE Transactions on Instrumentation and Measurement* 67.2 (Feb. 2018). Conference Name: IEEE Transactions on Instrumentation and Measurement, pp. 359–370. DOI: 10.1109/TIM.2017.2759398.

[12] Gabriel Tan et al. "Meta-Analysis of EEG Biofeedback in Treating Epilepsy". en. In: *Clinical EEG and Neuroscience* 40.3 (July 2009), pp. 173–179. DOI: 10.1177/155005940904000310.

[13] D. Corydon Hammond. "Neurofeedback Treatment of Depression and Anxiety". en. In: *Journal of Adult Development* 12.2-3 (Aug. 2005), pp. 131–137. DOI: 10.1007/s10804-005-7029-5.

[14] Robert T. Thibault, Michael Lifshitz, and Amir Raz. "The self-regulating brain and neurofeedback: Experimental science and clinical promise". en. In: *Cortex* 74 (Jan. 2016), pp. 247–261. DOI: 10.1016/j.cortex.2015.10.024.

[15] Aya Rezeika et al. "Brain–Computer Interface Spellers: A Review". en. In: (2018).

[16] Rajesh P. N. Rao. *Brain-computer interfacing : an introduction*. eng. New York : Cambridge University Press, 2013.

[17] Herbert Jasper and Wilder Penfield. "Electrocorticograms in man: Effect of voluntary movement upon the electrical activity of the precentral gyrus". en. In: *Archiv für Psychiatrie und Nervenkrankheiten* 183.1-2 (1949), pp. 163–174. DOI: 10.1007/BF01062488.

[18] J N Mak et al. "Optimizing the P300-based brain–computer interface: current status, limitations and future directions". en. In: *Journal of Neural Engineering* 8.2 (Mar. 2011), p. 025003. DOI: 10.1088/1741-2560/8/2/025003.

[19] Reza Fazel-Rezai et al. "P300 brain computer interface: current challenges and emerging trends". English. In: *Frontiers in Neuroengineering* 5 (July 2012). Publisher: Frontiers. DOI: 10.3389/fneng.2012.00014.

[20] Jon H Kaas and Pooja Balaram. "Current research on the organization and function of the visual system in primates". In: *Eye and Brain* 6.Suppl 1 (Sept. 2014), pp. 1–4. DOI: 10.2147/EB.S64016.

[21] Isabella C. Wagner, Ian Daly, and Aleksander Väljamäe. "Non-visual and Multisensory BCI Systems: Present and Future". en. In: *Towards Practical Brain-Computer Interfaces*. Ed. by Brendan Z. Allison et al. Series Title: Biological and Medical Physics, Biomedical Engineering. Berlin, Heidelberg: Springer Berlin Heidelberg, 2012, pp. 375–393. DOI: 10.1007/978-3-642-29746-5_19.

[22] Swati Banerjee and Viktor Jirsa. "Having sense of the path in cross modal correspondence in BCI studies: A comprehensive review of stimuli and EEG brain rhythms". en. In: *Work in progress* (2025).

[23] Wuyi Wang et al. "Spatio-temporal measures of electrophysiological correlates for behavioral multisensory enhancement during visual, auditory and somatosensory stimulation: A behavioral and ERP study". In: *Neuroscience Bulletin* 29.6 (Dec. 2013), pp. 715–724. DOI: 10.1007/s12264-013-1386-z.

[24] Marieke E. Thurlings et al. "Gaze-independent ERP-BCIs: augmenting performance through location-congruent bimodal stimuli". eng. In: *Frontiers in Systems Neuroscience* 8 (2014), p. 143. DOI: 10.3389/fnsys.2014.00143.

[25] Paul Bach-y-Rita. "Late postacute neurologic rehabilitation: neuroscience, engineering, and clinical programs". eng. In: *Archives of Physical Medicine and Rehabilitation* 84.8 (Aug. 2003), pp. 1100–1108. DOI: 10.1016/s0003-9993(03)00312-5.

[26] Chatrin Phunruangsakao et al. "Effects of visual-electrotactile stimulation feedback on brain functional connectivity during motor imagery practice". In: *Scientific Reports* 13 (Oct. 2023), p. 17752. DOI: 10.1038/s41598-023-44621-6.

[27] Carter Compton Collins. "Tactile Television - Mechanical and Electrical Image Projection". In: *IEEE Transactions on Man-Machine Systems* 11.1 (Mar. 1970). Conference Name: IEEE Transactions on Man-Machine Systems, pp. 65–71. DOI: 10.1109/TMMS.1970.299964.

[28] Milica Isaković et al. "Optimization of Semiautomated Calibration Algorithm of Multichannel Electrotactile Feedback for Myoelectric Hand Prosthesis". In: *Applied Bionics and Biomechanics* 2019 (Mar. 2019), p. 9298758. DOI: 10.1155/2019/9298758.

[29] Jian Dong et al. "Online Closed-Loop Control Using Tactile Feedback Delivered Through Surface and Subdermal Electrotactile Stimulation". eng. In: *Frontiers in Neuroscience* 15 (2021), p. 580385. DOI: 10.3389/fnins.2021.580385.

[30] Guohong Chai et al. "Electrotactile Feedback Improves Grip Force Control and Enables Object Stiffness Recognition While Using a Myoelectric Hand". eng. In: *IEEE transactions on neural systems and rehabilitation engineering: a publication of the IEEE Engineering in Medicine and Biology Society* 30 (2022), pp. 1310–1320. DOI: 10.1109/TNSRE.2022.3173329.

[31] Yichen Han et al. "Substitutive proprioception feedback of a prosthetic wrist by electrotactile stimulation". eng. In: *Frontiers in Neuroscience* 17 (2023), p. 1135687. DOI: 10.3389/fnins.2023.1135687.

[32] Semyoung Oh, Byung-Jun Yoon, and Hangue Park. "Location-based electrotactile feedback localizes hitting point in virtual-reality table tennis game". eng. In: *Biomedical Engineering Letters* 14.3 (May 2024), pp. 593–604. DOI: 10.1007/s13534-024-00354-7.

[33] Laura Schmitz et al. "Crossmodal correspondences as common ground for joint action". en. In: *Acta Psychologica* 212 (Jan. 2021), p. 103222. DOI: 10.1016/j.actpsy.2020.103222.

[34] Michael L. Lowe and Kelly L. Haws. "Sounds Big: The Effects of Acoustic Pitch on Product Perceptions". EN. In: *Journal of Marketing Research* 54.2 (Apr. 2017). Publisher: SAGE Publications Inc, pp. 331–346. DOI: 10.1509/jmr.14.0300.

[35] Barry E. Stein, Terrence R. Stanford, and Benjamin A. Rowland. "Multisensory Integration and the Society for Neuroscience: Then and Now". In: *The Journal of Neuroscience* 40.1 (Jan. 2020), pp. 3–11. DOI: 10.1523/JNEUROSCI.0737-19.2019.

[36] Daria Shumkova, Jacquir Sabir, and Swati Banerjee. "CrossModal Correspondence based MultisensoryIntegration: A pilot study showing how HAV cues can modulate the reaction time". en. In: *Submited* (2024).

[37] M. A. Meredith and B. E. Stein. "Spatial factors determine the activity of multisensory neurons in cat superior colliculus". eng. In: *Brain Research* 365.2 (Feb. 1986), pp. 350–354. DOI: 10.1016/0006-8993(86)91648-3.

[38] Ryan A. Stevenson et al. "Interactions between the spatial and temporal stimulus factors that influence multisensory integration in human performance". In: *Experimental brain research. Experimentelle Hirnforschung. Experimentation cerebrale* 219.1 (May 2012), pp. 121–137. DOI: 10.1007/s00221-012-3072-1.

[39] Ryan L. Miller et al. "Relative Unisensory Strength and Timing Predict Their Multisensory Product". en. In: *The Journal of Neuroscience* 35.13 (Apr. 2015), pp. 5213–5220. DOI: 10.1523/JNEUROSCI.4771-14.2015.

[40] PaulJ. Laurienti et al. "Semantic congruence is a critical factor in multisensory behavioral performance". en. In: *Experimental Brain Research* 158.4 (Oct. 2004). DOI: 10.1007/s00221-004-1913-2.

[41] Gabriel Pires et al. "Visuo-auditory stimuli with semantic, temporal and spatial congruence for a P300-based BCI: An exploratory test with an ALS patient in a completely locked-in state". eng. In: *Journal of Neuroscience Methods* 379 (Sept. 2022), p. 109661. DOI: 10.1016/j.jneumeth.2022.109661.

[42] Yi Lin and Hongwei Ding. "Multisensory Integration of Emotions in a Face-Prosody-Semantics Stroop Task". en. In: *NeuroManagement and Intelligent Computing Method on Multimodal Interaction*. Suzhou China: ACM, Oct. 2019, pp. 1–5. DOI: 10.1145/3357160.3357668.

[43] Shoko Kanaya, Kenji Kariya, and Waka Fujisaki. "Cross-Modal Correspondence Among Vision, Audition, and Touch in Natural Objects: An Investigation of the Perceptual Properties of Wood". EN. In: *Perception* 45.10 (Oct. 2016). Publisher: SAGE Publications Ltd STM, pp. 1099–1114. DOI: 10.1177/0301006616652018.

[44] Marieke E. Thurlings et al. "Does bimodal stimulus presentation increase ERP components usable in BCIs?" eng. In: *Journal of Neural Engineering* 9.4 (Aug. 2012), p. 045005. DOI: 10.1088/1741-2560/9/4/045005.

[45] Steven J. Luck. *An Introduction to the Event-Related Potential Technique, second edition*. en. Google-Books-ID: SzavAwAAQBAJ. MIT Press, May 2014.

[46] Emanuel Donchin. "Surprise!... Surprise?" en. In: *Psychophysiology* 18.5 (1981), pp. 493–513. DOI: 10.1111/j.1469-8986.1981.tb01815.x.

[47] Emanuel Donchin and Michael G. H. Coles. "Is the P300 component a manifestation of context updating?" en. In: *Behavioral and Brain Sciences* 11.03 (Sept. 1988), p. 357. DOI: 10.1017/S0140525X00058027.

[48] Rolf Verleger. "Event-related potentials and cognition: A critique of the context updating hypothesis and an alternative interpretation of P3". en. In: *Behavioral and Brain Sciences* 11.3 (Sept. 1988), pp. 343–356. DOI: 10.1017/S0140525X00058015.

[49] Aida Azlina Mansor, Salmi Mohd Isa, and Syaharudin Shah Mohd Noor. "P300 and decision-making in neuromarketing". en. In: *Neuroscience Research Notes* 4.3 (Sept. 2021). Number: 3, pp. 21–26. DOI: 10.31117/neuroscirn.v4i3.83.

[50] John Polich. "Updating P300: an integrative theory of P3a and P3b". eng. In: *Clinical Neurophysiology: Official Journal of the International Federation of Clinical Neurophysiology* 118.10 (Oct. 2007), pp. 2128–2148. DOI: 10.1016/j.clinph.2007.04.019.

[51] Hanhan Zhang et al. "Concentrate your mind by counting the flashing point: A new P300 pattern in BCI". In: *2016 Sixth International Conference on Information Science and Technology (ICIST)*. May 2016, pp. 36–41. DOI: 10.1109/ICIST.2016.7483382.

[52] J. Silva-Pereyra et al. "Delayed P300 during Sternberg and color discrimination tasks in poor readers". eng. In: *International Journal of Psychophysiology: Official Journal of the International Organization of Psychophysiology* 40.1 (Feb. 2001), pp. 17–32. DOI: 10.1016/s0167-8760(00)00123-9.

[53] M. Fabiani, D. Karis, and E. Donchin. "Effects of mnemonic strategy manipulation in a Von Restorff paradigm". eng. In: *Electroencephalography and Clinical Neurophysiology* 75.2 (Feb. 1990), pp. 22–35. DOI: 10.1016/0013-4694(90)90149-e.

[54] Yun Chen et al. "Proactive and reactive control differ between task switching and response rule switching: Event-related potential evidence". en. In: *Neuropsychologia* 172 (July 2022), p. 108272. DOI: 10.1016/j.neuropsychologia.2022.108272.

[55] Yuqin Deng et al. "Conflict monitoring and adjustment in the task-switching paradigm under different memory load conditions: an ERP/sLORETA analysis". en. In: (2015).

[56] Richard Simson, Herbert G Vaughan, and Walter Ritter. "The scalp topography of potentials in auditory and visual Go/NoGo tasks". en. In: *Electroencephalography and Clinical Neurophysiology* 43.6 (Dec. 1977), pp. 864–875. DOI: 10.1016/0013-4694(77)90009-8.

[57] Jun'ichi Katayama and John Polich. "Auditory and visual P300 topography from a 3 stimulus paradigm". en. In: *Clinical Neurophysiology* 110.3 (Mar. 1999), pp. 463–468. DOI: 10.1016/S1388-2457(98)00035-2.

[58] Mathis Fleury et al. "A Survey on the Use of Haptic Feedback for Brain-Computer Interfaces and Neurofeedback". English. In: *Frontiers in Neuroscience* 14 (June 2020). Publisher: Frontiers. DOI: 10.3389/fnins.2020.00528.

[59] G.R. Muller-Putz et al. "Steady-state somatosensory evoked potentials: suitable brain signals for brain-computer interfaces?" In: *IEEE Transactions on Neural Systems and Rehabilitation Engineering* 14.1 (Mar. 2006). Conference Name: IEEE Transactions on Neural Systems and Rehabilitation Engineering, pp. 30–37. DOI: 10.1109/TNSRE.2005.863842.

[60] Christian Breitwieser et al. "Somatosensory evoked potentials elicited by stimulating two fingers from one hand– usable for BCI?" eng. In: *Annual International Conference of the IEEE Engineering in Medicine and Biology Society. IEEE Engineering in Medicine and Biology Society. Annual International Conference* 2011 (2011), pp. 6373–6376. DOI: 10.1109/IEMBS.2011.6091573.

[61] Sangtae Ahn et al. "Achieving a hybrid brain–computer interface with tactile selective attention and motor imagery". en. In: *Journal of Neural Engineering* 11.6 (Oct. 2014). Publisher: IOP Publishing, p. 066004. DOI: 10.1088/1741-2560/11/6/066004.

[62] Anthony M. Norcia et al. "The steady-state visual evoked potential in vision research: A review". In: *Journal of Vision* 15.6 (May 2015), p. 4. DOI: 10.1167/15.6.4.

[63] Gary Garcia-Molina and Danhua Zhu. "Optimal spatial filtering for the steady state visual evoked potential: BCI application". In: *2011 5th International IEEE/EMBS Conference on Neural Engineering*. ISSN: 1948-3554. Apr. 2011, pp. 156–160. DOI: 10.1109/NER.2011.5910512.

[64] Minglun Li et al. "Brain–Computer Interface Speller Based on Steady-State Visual Evoked Potential: A Review Focusing on the Stimulus Paradigm and Performance". en. In: *Brain Sciences* 11.4 (Apr. 2021). Number: 4 Publisher: Multidisciplinary Digital Publishing Institute, p. 450. DOI: 10.3390/brainsci11040450.

[65] Rafał Kuś et al. "Integrated trimodal SSEP experimental setup for visual, auditory and tactile stimulation". eng. In: *Journal of Neural Engineering* 14.6 (Dec. 2017), p. 066002. DOI: 10.1088/1741-2552/aa836f.

[66] Nienke van Atteveldt et al. "Multisensory Integration: Flexible Use of General Operations". en. In: *Neuron* 81.6 (Mar. 2014), pp. 1240–1253. DOI: 10.1016/j.neuron.2014.02.044.

[67] Dora E Angelaki, Yong Gu, and Gregory C DeAngelis. "Multisensory integration: psychophysics, neurophysiology, and computation". en. In: *Current Opinion in Neurobiology* 19.4 (Aug. 2009), pp. 452–458. DOI: 10.1016/j.conb.2009.06.008.

[68] G. Vinodh Kumar et al. "Large Scale Functional Brain Networks Underlying Temporal Integration of Audio-Visual Speech Perception: An EEG Study". en. In: *Frontiers in Psychology* 7 (Oct. 2016). DOI: 10.3389/fpsyg.2016.01558.

[69] Pietro Aricò et al. "Adaptive Automation Triggered by EEG-Based Mental Workload Index: A Passive Brain-Computer Interface Application in Realistic Air Traffic Control Environment". en. In: *Frontiers in Human Neuroscience* 10 (Oct. 2016). DOI: 10.3389/fnhum.2016.00539.

[70] Hilit Serby, Elad Yom-Tov, and Gideon F. Inbar. "An improved P300-based brain-computer interface". eng. In: *IEEE transactions on neural systems and rehabilitation engineering: a publication of the IEEE Engineering in Medicine and Biology Society* 13.1 (Mar. 2005), pp. 89–98. DOI: 10.1109/TNSRE.2004.841878.

[71] Brenda K Wiederhold et al. "An Investigation into Physiological Responses in Virtual Environments: An Objective Measurement of Presence". en. In: *Towards Cyberpsychology: Mind, Cognition, and Society in the Internet Age*. IOS Press, 2001, pp. 175–184.

[72] Sandra G. Hart and Lowell E. Staveland. "Development of NASA-TLX (Task Load Index): Results of empirical and theoretical research". In: *Human mental workload*. Advances in psychology, 52. Oxford, England: North-Holland, 1988, pp. 139–183. DOI: 10.1016/S0166-4115(08)62386-9.

[73] Abdulaziz Alshaer, Holger Regenbrecht, and David O'Hare. "Immersion factors affecting perception and behaviour in a virtual reality power wheelchair simulator". en. In: *Applied Ergonomics* 58 (Jan. 2017), pp. 1–12. DOI: 10.1016/j.apergo.2016.05.003.

[74] A. Lecuyer et al. "Brain-Computer Interfaces, Virtual Reality, and Videogames". en. In: *Computer* 41.10 (Oct. 2008), pp. 66–72. DOI: 10.1109/MC.2008.410.

[75] Reza Abbasi-Asl, Mohammad Keshavarzi, and Dorian Yao Chan. "Brain-Computer Interface in Virtual Reality". en. In: *2019 9th International IEEE/EMBS Conference on Neural Engineering (NER)*. arXiv:1811.06040 [cs]. Mar. 2019, pp. 1220–1224. DOI: 10.1109/NER.2019.8717158.

[76] Mikko Kytö, Kenta Kusumoto, and Pirkko Oittinen. "The Ventriloquist Effect in Augmented Reality". In: *2015 IEEE International Symposium on Mixed and Augmented Reality*. Sept. 2015, pp. 49–53. DOI: 10.1109/ISMAR.2015.18.

[77] James Wen et al. "Really, it's for your own good...making augmented reality navigation tools harder to use". en. In: *CHI '14 Extended Abstracts on Human Factors in Computing Systems*. Toronto Ontario Canada: ACM, Apr. 2014, pp. 1297–1302. DOI: 10.1145/2559206.2581156.

[78] Ernst Kruijff, J. Edward Swan, and Steven Feiner. "Perceptual issues in augmented reality revisited". In: *2010 IEEE International Symposium on Mixed and Augmented Reality*. Oct. 2010, pp. 3–12. DOI: 10.1109/ISMAR.2010.5643530.

[79] Patricia Cornelio, Carlos Velasco, and Marianna Obrist. "Multisensory Integration as per Technological Advances: A Review". en. In: *Frontiers in Neuroscience* 15 (June 2021), p. 652611. DOI: 10.3389/fnins.2021.652611.

[80] Christopher J. Wilson and Alessandro Soranzo. "The Use of Virtual Reality in Psychology: A Case Study in Visual Perception". en. In: *Computational and Mathematical Methods in Medicine* 2015.1 (2015). _eprint: https://onlinelibrary.wiley.com/doi/pdf/10.1155/2015/151702, p. 151702. DOI: 10.1155/2015/151702.

[81] Maria Gallagher and Elisa Raffaella Ferrè. "Cybersickness: a Multisensory Integration Perspective". en. In: (Jan. 2018). Publisher: Brill. DOI: 10.1163/22134808-20181293.

[82] Ryuji Hirayama et al. "A volumetric display for visual, tactile and audio presentation using acoustic trapping". en. In: *Nature* 575.7782 (Nov. 2019). Publisher: Nature Publishing Group, pp. 320–323. DOI: 10.1038/s41586-019-1739-5.

[83] Sophie Deneve and Alexandre Pouget. "Bayesian multisensory integration and cross-modal spatial links". en. In: *Journal of Physiology-Paris* 98.1-3 (Jan. 2004), pp. 249–258. DOI: 10.1016/j.jphysparis.2004.03.011.

[84] Diego Martinez Plasencia et al. "GS-PAT: high-speed multi-point sound-fields for phased arrays of transducers". en. In: *ACM Transactions on Graphics* 39.4 (Aug. 2020). DOI: 10.1145/3386569.3392492.

[85] Valeria Bruschi et al. "A Review on Head-Related Transfer Function Generation for Spatial Audio". en. In: *Applied Sciences* 14.23 (Jan. 2024). Number: 23 Publisher: Multidisciplinary Digital Publishing Institute, p. 11242. DOI: 10.3390/app142311242.

[86] Michał Pec, Michał Bujacz, and Paweł Strumiłło. "Head related transfer functions measurement and processing for the purpose of creating a spatial sound environment". In: *Photonics Applications in Astronomy, Communications, Industry, and High-Energy Physics Experiments 2007*. Vol. 6937. SPIE, Dec. 2007, pp. 966–973. DOI: 10.1117/12.784835.

[87] Gianluca Memoli et al. "Metamaterial bricks and quantization of meta-surfaces". en. In: *Nature Communications* 8.1 (Feb. 2017). Publisher: Nature Publishing Group, p. 14608. DOI: 10.1038/ncomms14608.

[88] Mohd Adili Norasikin et al. "SoundBender: Dynamic Acoustic Control Behind Obstacles". In: *Proceedings of the 31st Annual ACM Symposium on User Interface Software and Technology*. UIST '18. New York, NY, USA: Association for Computing Machinery, Oct. 2018, pp. 247–259. DOI: 10.1145/3242587.3242590.

[89] J.J. Yuan et al. "The valence strength of negative stimuli modulates visual novelty processing: Electrophysiological evidence from an event-related potential study". en. In: *Neuroscience* 157.3 (Dec. 2008), pp. 524–531. DOI: 10.1016/j.neuroscience.2008.09.023.

[90]   Eric Halgren, Ksenija Marinkovic, and Patrick Chauvel. "Generators of the late cognitive potentials in auditory and visual oddball tasks". en. In: *Electroencephalography and Clinical Neurophysiology* 106.2 (Feb. 1998), pp. 156–164. DOI: 10.1016/S0013-4694(97)00119-3.

[91]   Rui Hu et al. "The Neural Responses of Visual Complexity in the Oddball Paradigm: An ERP Study". In: *Brain Sciences* 12.4 (Mar. 2022), p. 447. DOI: 10.3390/brainsci12040447.

[92]   Elena V. Orekhova et al. "Frequency of gamma oscillations in humans is modulated by velocity of visual motion". en. In: *Journal of Neurophysiology* 114.1 (July 2015), pp. 244–255. DOI: 10.1152/jn.00232.2015.

[93]   M. Krebber et al. "Visuotactile motion congruence enhances gamma-band activity in visual and somatosensory cortices". In: *NeuroImage* 117 (2015), pp. 160–169. DOI: 10.1016/j.neuroimage.2015.05.056.

[94]   Jonas Misselhorn et al. "A matter of attention: Crossmodal congruence enhances and impairs performance in a novel trimodal matching paradigm". en. In: *Neuropsychologia* 88 (July 2016), pp. 113–122. DOI: 10.1016/j.neuropsychologia.2015.07.022.

[95]   Kyuto Uno and Kazuhiko Yokosawa. "Cross-modal correspondence between auditory pitch and visual elevation modulates audiovisual temporal recalibration". en. In: *Scientific Reports* 12.1 (Dec. 2022). Publisher: Nature Publishing Group, p. 21308. DOI: 10.1038/s41598-022-25614-3.

[96]   Sylvie Nozaradan, Isabelle Peretz, and André Mouraux. "Steady-state evoked potentials as an index of multisensory temporal binding". en. In: *NeuroImage* 60.1 (Mar. 2012), pp. 21–28. DOI: 10.1016/j.neuroimage.2011.11.065.

[97]   M. Binder. "Neural correlates of audiovisual temporal processing – Comparison of temporal order and simultaneity judgments". In: *Neuroscience* 300 (Aug. 2015), pp. 432–447. DOI: 10.1016/j.neuroscience.2015.05.011.

[98]   Patrick Bruns. "The Ventriloquist Illusion as a Tool to Study Multisensory Processing: An Update". en. In: *Frontiers in Integrative Neuroscience* 13 (Sept. 2019), p. 51. DOI: 10.3389/fnint.2019.00051.

[99]   Souta Hidaka, Wataru Teramoto, and Yoichi Sugita. "Spatiotemporal Processing in Crossmodal Interactions for Perception of the External World: A Review". en. In: *Frontiers in Integrative Neuroscience* 9 (Dec. 2015). DOI: 10.3389/fnint.2015.00062.

[100]  Anne Caclin et al. "Tactile "capture" of audition". en. In: *Perception & Psychophysics* 64.4 (May 2002), pp. 616–630. DOI: 10.3758/BF03194730.

[101]  Majed Samad and Ladan Shams. "Recalibrating the body: visuotactile ventriloquism aftereffect". en. In: *PeerJ* 6 (Mar. 2018). Publisher: PeerJ Inc., e4504. DOI: 10.7717/peerj.4504.

[102]  Kaisa Tiippana. "What is the McGurk effect?" en. In: *Frontiers in Psychology* 5 (July 2014). DOI: 10.3389/fpsyg.2014.00725.

[103]  Jeffrey M. Yau, Alison I. Weber, and Sliman J. Bensmaia. "Separate Mechanisms for Audio-Tactile Pitch and Loudness Interactions". en. In: *Frontiers in Psychology* 1 (2010). DOI: 10.3389/fpsyg.2010.00160.

[104]  M.P.M. Kammers et al. "The rubber hand illusion in action". en. In: *Neuropsychologia* 47.1 (Jan. 2009), pp. 204–211. DOI: 10.1016/j.neuropsychologia.2008.07.028.

[105]  Micah M. Murray et al. "Multisensory Processes: A Balancing Act across the Lifespan". en. In: *Trends in Neurosciences* 39.8 (Aug. 2016), pp. 567–579. DOI: 10.1016/j.tins.2016.05.003.

[106]  M Falkenstein, J Hoormann, and J Hohnsbein. "ERP components in Go/Nogo tasks and their relation to inhibition". In: *Acta Psychologica* 101.2 (Apr. 1999), pp. 267–291. DOI: 10.1016/S0001-6918(99)00008-6.

[107]  R.j. Huster et al. "The role of the cingulate cortex as neural generator of the N200 and P300 in a tactile response inhibition task". en. In: *Human Brain Mapping* 31.8 (2010). _eprint: https://onlinelibrary.wiley.com/doi/pdf/10.1002/hbm.20933 pp. 1260–1271. DOI: 10.1002/hbm.20933.